\documentclass[12pt,aps,pra,reprint,superscriptaddress,amsmath,amsfonts,amssymb,longbibliography]{revtex4-1}
\usepackage[]{revquantum}
\usepackage[english]{babel}
\usepackage{mathrsfs}
\usepackage{hyperref}
\usepackage{hypernat}
\usepackage[utf8]{inputenc}
\usepackage[caption=false]{subfig}
\usepackage{graphicx}
\usepackage{xcolor}
\begin{document}
\title{Diagnosing quantum chaos with out-of-time-ordered-correlator quasiprobability \\
in the kicked-top model}
\author{José Raúl González Alonso}
\affiliation{Schmid College of Science and Technology, Chapman University, Orange, CA 92866, USA}
\author{Nathan Shammah}
\affiliation{Unitary Fund}
\affiliation{Theoretical Quantum Physics Laboratory, RIKEN Cluster for Pioneering Research, Wako-shi, Saitama 351-0198, Japan}
\affiliation{Quantum Technology Lab, Dipartimento di Fisica, Universit\`{a} degli Studi di Milano, 20133 Milano, Italy}
\author{Shahnawaz Ahmed}
\affiliation{Theoretical Quantum Physics Laboratory, RIKEN Cluster for Pioneering Research, Wako-shi, Saitama 351-0198, Japan}
\affiliation{Department of Microtechnology and Nanoscience, Applied Quantum Physics Laboratory,  Chalmers University of Technology, Göteborg, Sweden.}
\author{Franco Nori}
\affiliation{Theoretical Quantum Physics Laboratory, RIKEN Cluster for Pioneering Research, Wako-shi, Saitama 351-0198, Japan}
\affiliation{RIKEN Center for Quantum Computing (RQC), Wako-shi, Saitama 351-0198, Japan}

\affiliation{Department of Physics, University of Michigan, Ann Arbor, Michigan 48109-1040, USA}
\author{Justin Dressel}
\affiliation{Schmid College of Science and Technology, Chapman University, Orange, CA 92866, USA}
\affiliation{Institute for Quantum Studies, Chapman University, Orange, CA 92866, USA}

\date{\today}

\begin{abstract}
While classical chaos has been successfully characterized with consistent theories and intuitive techniques, such as with the use of Lyapunov exponents, quantum chaos is still poorly understood, as well as its relation with multi-partite entanglement and information scrambling. We consider a benchmark system, the kicked top model, which displays chaotic behaviour in the classical version, and proceed to characterize the quantum case with a thorough diagnosis of the growth of chaos and entanglement in time. As a novel tool for the characterization of quantum chaos, we introduce for this scope the \emph{quasi-probability distribution} behind the out-of-time-ordered correlator (OTOC). 
 We calculate the cumulative nonclassicality of this distribution, which has already been shown to outperform the simple use of OTOC as a probe to distinguish between integrable and nonintegrable Hamiltonians. To provide a thorough comparative analysis, we contrast the behavior of the nonclassicality with entanglement measures, such as the tripartite mutual information of the Hamiltonian as well as the entanglement entropy. We find that systems whose initial states would lie in the ``sea of chaos" in the classical kicked-top model, exhibit, as they evolve in time, characteristics associated with chaotic behavior and entanglement production in closed quantum systems. We corroborate this indication by capturing it with this novel OTOC-based measure. 
\end{abstract}

\maketitle

\section{Introduction}
The concept of scrambling, initially introduced to characterize quantum chaos, has since been applied to quantum information processing, as it quantifies the delocalization of quantum information in a system.
Out-of-time-ordered correlators (OTOCs) have been the subject of intense research as witnesses of scrambling, and as tools to study entanglement dynamics and quantum chaos \cite{Larkin_QuasiclassicalMethodTheory_1969,Kitaev_HiddenCorrelationsHawking_2014,Shenker_BlackHolesButterfly_2014,Shenker_MultipleShocks_2014,Hartnoll_TheoryUniversalIncoherent_2015,Shenker_StringyEffectsScrambling_2015,Kitaev_SimpleModelQuantum_2015,Roberts_LocalizedShocks_2015,Roberts_DiagnosingChaosUsing_2015,Maldacena_BoundChaos_2016,Aleiner_MicroscopicModelQuantum_2016,Blake_UniversalChargeDiffusion_2016,Blake_UniversalDiffusionIncoherent_2016,Chen_UniversalLogarithmicScrambling_2016,Lucas_ChargeDiffusionButterfly_2016,Roberts_LiebRobinsonBoundButterfly_2016,Hosur_ChaosQuantumChannels_2016,Banerjee_SolvableModelDynamical_2017,Fan_OutoftimeorderCorrelationManybody_2017,Gu_LocalCriticalityDiffusion_2017,Roberts_ChaosComplexityDesign_2017,Chen_OperatorScramblingQuantum_2018,Huang_OutoftimeorderedCorrelatorsManybody_2017,Iyoda_ScramblingQuantumInformation_2018,Yoshida_EfficientDecodingHaydenPreskill_2017,Lin_OutoftimeorderedCorrelatorsQuantum_2018,Pappalardi_ScramblingEntanglementSpreading_2018,YungerHalpern_EntropicUncertaintyRelations_2019,Vermersch_2019, Harrow21}. However, a true understanding of the nature of quantum chaos and the limits of the usefulness of various diagnostic tools such as OTOCs to study chaos are the source of ongoing investigations both theoretically~\cite{Cao2021}, and in recent experiments~\cite{Braumller2021, Mi2021}.

As OTOCs measure time correlations among initially commuting operators, they provide a quantum counterpart to the classical and semiclassical theory of characterizing chaos using diverging trajectories with Lyapunov exponents. 

While there have been indications of a connection between quantum chaos and scrambling, their mutual relationship needs to be fully elucidated, as well as their relationship with entanglement: for example, it has been shown that OTOCs can characterize chaotic dynamics in the quantum regime even where current entanglement witnesses saturate \cite{Pappalardi_ScramblingEntanglementSpreading_2018}. An analytical relationship between OTOCs and entanglement entropy, which is related to classical chaos quantifiers, has been shown in Ref. \cite{Lerose20}. In the semiclassical regime this predicts the usual exponential growth for the OTOC and a linear growth of the entanglement entropy, characterized by the same Lyapunov exponent.

The cumulative nonclassicality of the quasi-probability distribution behind the out-of-time-ordered correlator exhibits 
different time scales that have been conjectured to be useful for distinguishing integrable and nonintegrable Hamiltonians 
\cite{GonzalezAlonso_OutofTimeOrderedCorrelatorQuasiprobabilitiesRobustly_2019}. That is, the time scales corresponding to regions where 
the quasi-probability distribution becomes negative or has a nonzero imaginary part can be used to identify behavior that cannot be explained classically. However, this conjecture was proposed in the context of spin chains with a notion of spatial locality \cite{GonzalezAlonso_OutofTimeOrderedCorrelatorQuasiprobabilitiesRobustly_2019}. 

In this work, we use the cumulative nonclassicality of the quasi-probability distribution to better characterize quantum information scrambling and quantum chaos in the quantum kicked top. Since this model contains second-momenta of collective angular momentum operators, which induce long-range interactions, it is a good candidate to study the interplay of quantum chaos and entanglement in a many-body quantum system. Moreover, since the simple kicked-top model has been extensively studied, e.g., the chaotic behavior is well understood in its classical counterpart, it represents a natural target to test and compare our novel analytical tool. 

Additionally, we contrast the behavior of both the OTOC and the cumulative nonclassicality of its quasi-probability distribution with other measures used to diagnose scrambling and chaos. In particular, we study the von Neumann entropy of entanglement for a one to many partition of the system 
and the tripartite mutual information (TMI)\cite{Cerf_InformationTheoryQuantum_1998,Hosur_ChaosQuantumChannels_2016} of the unitary channel generated by the kicked-top Hamiltonian to elucidate the relationship between quantum chaos and
scrambling in many-body systems that lack a notion of spatial locality.

We find that for small system sizes, all of the considered measures of quantum chaos and scrambling give poor, and even misleading, diagnostics. We provide an interpretation for these results, tailoring the standard notion of scrambling, based on the delocalization of information, to the case of the kicked-top, which lacks a notion of spatial locality.

This paper is organized as follows. We begin in Sec.~\ref{sec:kicked-top} by presenting the model of a kicked-top that we use in 
our studies and present its classical phase space with chaotic features. In Sec.~\ref{sec:otoc-and-qc} we motivate the study of OTOCs by using the square of the commutator between two 
operators. Then, in Sec.~\ref{sec:measures}, we introduce the various measures of quantum chaos and entanglement that we will compare:  in Sec.~\ref{sec:otoc-qp-nc} we define the (coarse-grained) quasiprobability behind the OTOC and introduce a measure of nonclassicality based on it. In Sec.~\ref{sec:TMI} we present the tripartite mutual information definition for a channel and explain how it relates to a more general notion of scrambling. Our numerical results are shown in Sec.~\ref{sec:numerics-results}, where the various measures are compared both in time and in conjugate space, while the details of our calculations are presented in Appendix \ref{sec:sim-details}. Finally, in Sec.~\ref{sec:conclusions} we present our conclusions and discuss future outlook of this work.

\section{Quantum kicked-top model and the semiclassical limit}
\label{sec:kicked-top}
We are interested in studying the OTOC, and its coarse-grained quasiprobability distribution, for a system without a notion of spatial 
locality and that can be considered chaotic and scrambling. For our purposes, we will consider a system evolving under the \emph{
kicked-top} Hamiltonian. Following Ref.~\cite{Wang_EntanglementSignatureQuantum_2004}, we consider the Hamiltonian 
(setting $\hbar=1$), 
\begin{align}\label{eqn:qkt-hamilt}
H =  \frac{\kappa}{2j\tau}J_z^2 + pJ_y\sum_{n=-\infty}^{\infty}\delta\left(\frac{t}{\tau}-n_k\right).
\end{align}
In Eq.~\eqref{eqn:qkt-hamilt}, $\tau$ represents the duration between periodic kicks (as indicated by the presence of the
Dirac delta function), $p$ is the strength of each kick (i.e. a turn by an angle $p$ by each kick), $\kappa$ is the
(dimensionless) strength of the twist and $\omega_0$ is the frequency constant. The twists are represented by the $J_z^2$ term whereas turns are associated with
the $J_y$ term. Additionally, the operators $J_\alpha,\; \alpha\in\{x,y,z\}$ denote collective spin operators. These
operators can be chosen to describe a system of $N$ spins such that if we represent the Pauli 
operators for the $i$-th spin by $\sigma_\alpha^i$ then
\begin{align}
J_\alpha = \frac{1}{2}\sum_{i=1}^{N} \sigma^i_\alpha.
\end{align}
With the above definition the value $j$ of the collective angular momentum satisfies $j=N/2$. Furthermore, we will use the following
conventions
\begin{subequations}
\begin{align}
U_{\text{kick}} := \exp\left(-\mathrm{i}p J_y\right) \\
U_{\text{twist}}(t) := \exp\left(-\mathrm{i} \frac{\kappa}{2j\tau} J_z^2 t\right).
\end{align}
\end{subequations}

In the calculations below we use the parameters $\kappa=3.0$, $p=\pi/2$, and $\tau=1.0$ since they allow us to study different kinds of
behaviors in phase space \cite{Wang_EntanglementSignatureQuantum_2004}. Additionally, for our initial states, and to be as close to a classical state as possible, we will use the spin coherent states \cite{P.W.Atkins_AngularMomentumCoherent_1971,Arecchi_AtomicCoherentStates_1972,Zhang_CoherentStatesTheory_1990} 
given by
\begin{align}\label{eqn:cohenrent_state_def}
\ket{\theta,\phi} = \exp\left[(-\ii \theta (\sin(\phi) J_x - \cos(\phi) J_y))\right] \ket{j,-j},
\end{align}
where $\ket{j,-j}$ represents the lowest value eigenstate of the $J_z$ operator.

\begin{figure}[!htbp]
    \includegraphics[width=1.0\columnwidth]{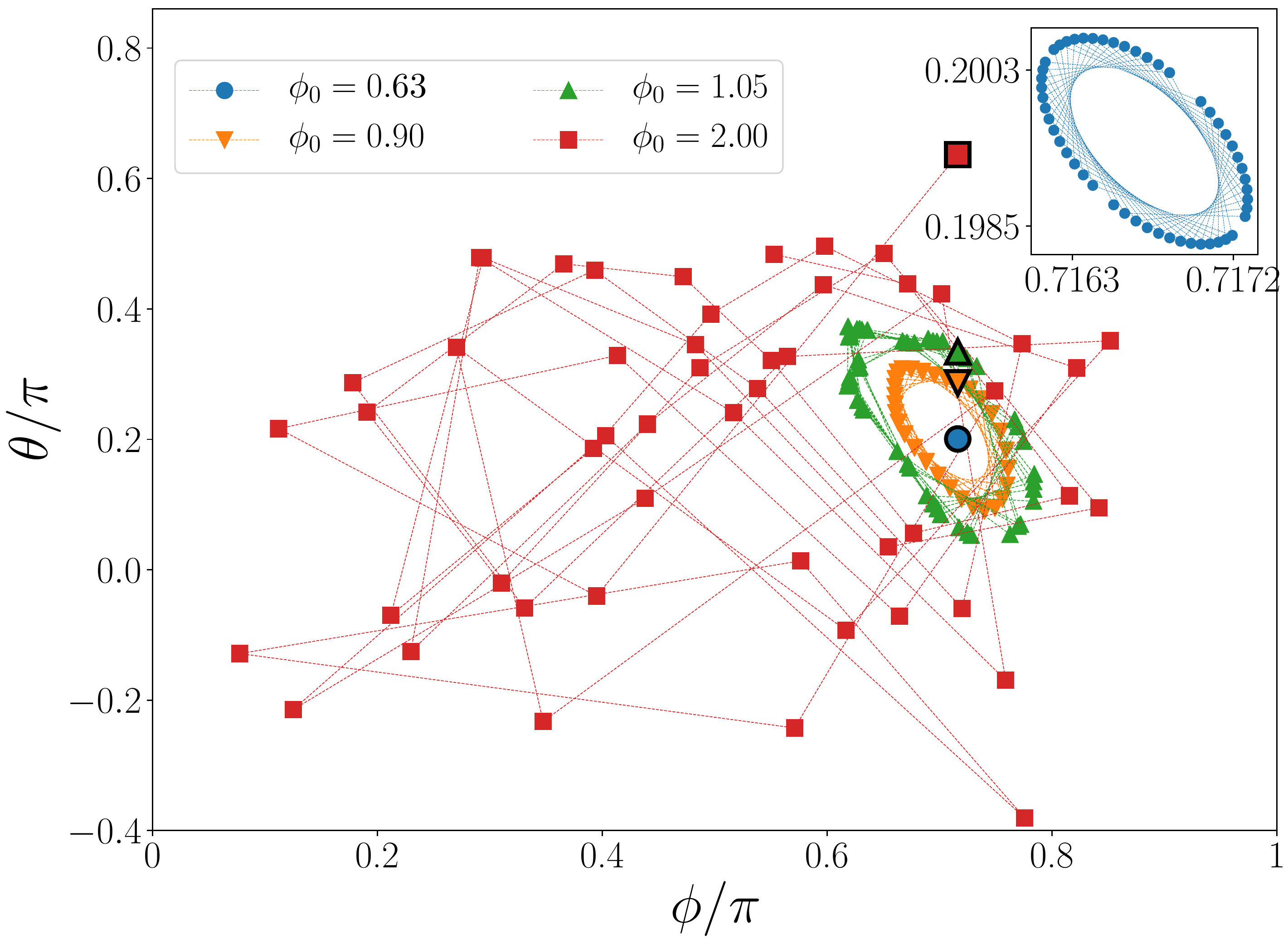}
    \caption{(Color online) Illustration of the dynamics of the classical kicked-top with $\kappa=3.0$, $p=\pi/2$,
    and $\tau=1.0$ for several initial conditions and $n_k=50$ kicks. We use the following initial conditions:
    (a) \emph{elliptic fixed point} with $\theta_0 = 2.25$, $\phi_0 = 0.63$, (b) \emph{regular region point}
    with $\theta_0 = 2.25$, $\phi_0 = 0.90$, (c) \emph{edge of chaos point} with $\theta_0 = 2.25$, $\phi_0 = 1.05$, and
    (d) \emph{sea of chaos point} with $\theta_0 = 2.25$, $\phi_0 = 2.0$. The markers relative to the four points of phase space for the initial conditions are larger marker and have a black edge. Inset: the area of the elliptic fixed point is magnified.
    \label{fig:stroboscopic_map_nk_e_50}}
\end{figure}

In order to sharpen our intuition, we first consider the case below the semiclassical limit. That is, in the limit in which $j\rightarrow\infty$, we define $X=\braket{J_x}/j$, $Y=\braket{J_y}/j$, and $Z=\braket{J_z}/j$. Under such conditions, it is possible to obtain the following equations of motion for the special case in which $p=\pi/2$ \cite{Haake_ClassicalQuantumChaos_1987,Ghose_QuantumChaosTunneling_2008,Ghose_ChaosEntanglementDecoherence_2008,Kumari_UntanglingEntanglementChaos_2019,Haake_QuantumSignaturesChaos_2001},
\begin{subequations}
\begin{align}
X ((n+1)\tau) &= Z (n\tau)\cos(\kappa X(n\tau)) + Y(n\tau)\sin(\kappa X(n\tau)),\\
Y ((n+1)\tau) &= -Z (n\tau)\sin(\kappa X(n\tau)) + Y(n\tau)\cos(\kappa X(n\tau)),\\
Z ((n+1)\tau) &= -X(n\tau)
\end{align}
\end{subequations}
where $X(n)$, $Y(n)$, and $Z(n)$ denote the values of $X$, $Y$, and $Z$ after $n$ applications of the evolution operator $ U_{\text{kick}} U_{\text{twist}}(\tau)$. In Fig.~\ref{fig:stroboscopic_map_nk_e_50}, we show a simple stroboscopic plot of the different initial coherent states that we use in our calculations 
and how they traverse the phase space for a total of $n_k=50$ kicks.
As we can appreciate in Fig.~\ref{fig:stroboscopic_map_nk_e_50}, as we get closer to the region of the phase space associated with chaos, 
the motion becomes more erratic and covers a wider section. This is shown by fixing $\theta_0$ and choosing several initial conditions for $\phi_0$: for $\phi_0 = 0.63$ (circles), an elliptic fixed point is found; at $\phi_0 = 2.0$, there is a regular region (triangles), which reaches an edge at $\phi_0 = 1.05$, while for $\phi_0 = 2.0$ the system evolves in a sea of chaos.

In contrast to the semiclassical limit, when we study the purely quantum case, we can no longer describe states in phase space by 
points. Instead, we have to make use of distributions. While there are different possible distributions that can be used,
for simplicity, we will use the Husimi $Q$ function \cite{Husimi_FormalPropertiesDensity_1940} to illustrate the behavior of the 
kicked-top in the quantum case. We define the Husimi $Q$ function as
\begin{align}
Q_\rho(\theta,\phi) := \braket{\theta,\phi|\rho|\theta,\phi},
\end{align}
with $\rho$ as the density matrix of the physical state of the system and $\ket{\theta,\phi}$ given by Eq.~\eqref{eqn:cohenrent_state_def}.

It is not difficult to see that for our choice of initial state, that is, a coherent state $\ket{\theta_0,\phi_0}$, its $Q$ function 
is given by
\begin{align}\label{eqn:Q_func_coh_states}
Q_{\ket{\theta_0,\phi_0}}(\theta,\phi) = \left(\cos\frac{\Theta}{2}\right)^{4j},
\end{align}
where $\Theta$ is the angle such that $\cos \Theta = \cos \theta\cos\theta_0 + \sin\theta\sin\theta_0 \left( 
\cos\phi\cos\phi_0 + \sin\phi\sin\phi_0 \right)$. In other words, $\Theta$ is the angle between the vectors of the two points
$(\theta,\phi)$ and $(\theta_0,\phi_0)$ on the unit sphere.

\begin{figure*}[!htbp]
    \includegraphics[width=2\columnwidth]{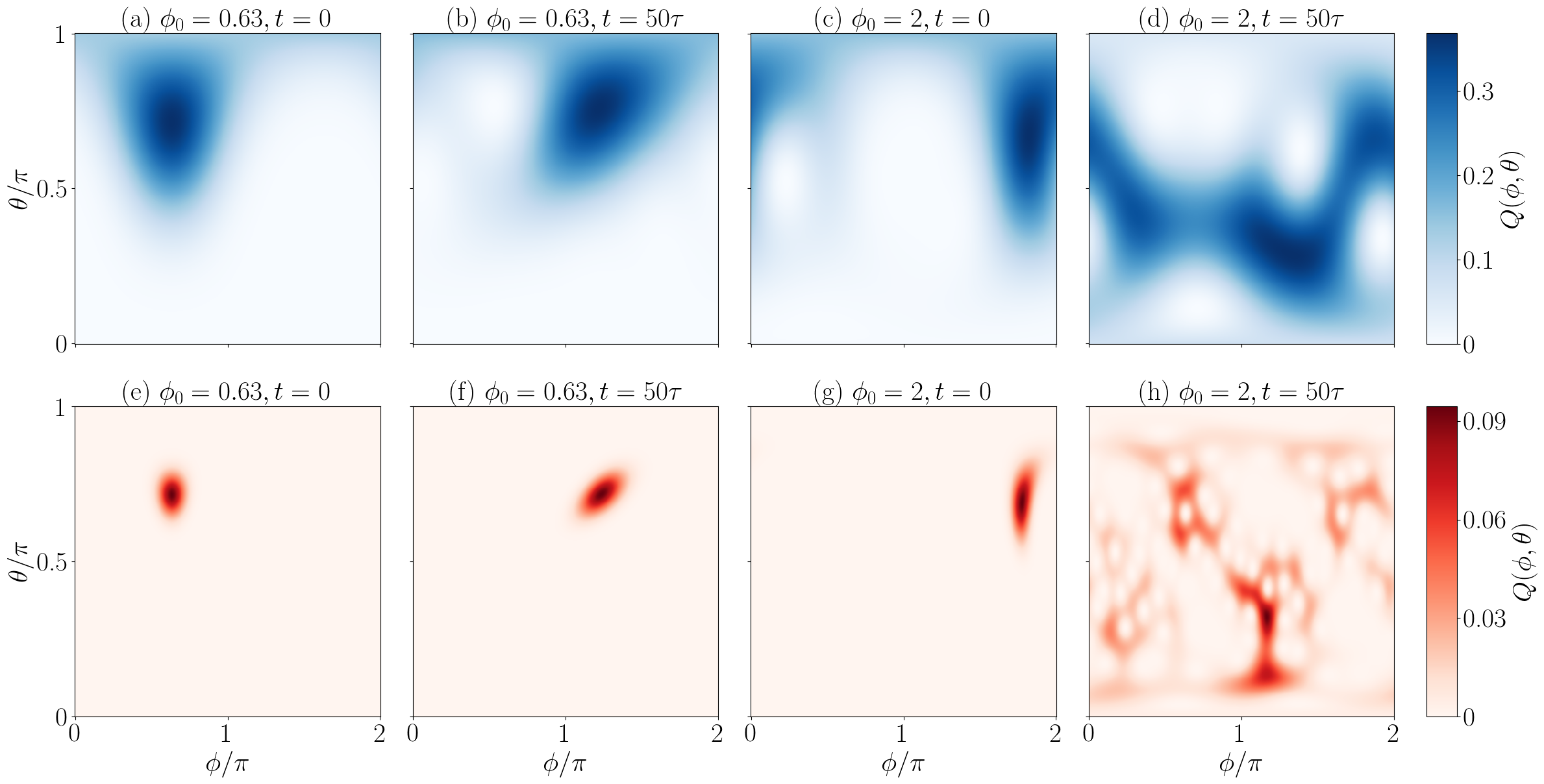}
    \caption{(Color online) 
    Contour plot of the Husimi function $Q(\theta,\phi)$ for a system evolving under the Hamiltonian of Eq.~(\ref{eqn:qkt-hamilt}), with $\kappa=3.0$, $p=\pi/2$, and $\tau=1.0$. Two systems are considered,  $N=5$ spins (top row) and $N=100$ spins (bottom row), for which $t=0$ is shown in panels (a), (c), (e), (g), and at time $t=50\tau$, in panels (b), (d), (f), (h), for two initial conditions -- $\theta_0=2.25$, $\phi_0=0.63$ and $\theta_0=2.25$, $\phi_0=2$, corresponding to the elliptic fixed point  and the sea of chaos in Fig.~\ref{fig:stroboscopic_map_nk_e_50}.}
    \label{fig:Husimi_comparison_N_5}
\end{figure*}

It is clear from Eq.~\eqref{eqn:Q_func_coh_states} that as the number of spins increases, the distribution of our initial state becomes 
narrower. We illustrate this behavior in Fig.~\ref{fig:Husimi_comparison_N_5} for 5 spins (top row) and 100 spins (bottom row), respectively, using the same initial points in the phase space as those shown in Fig. 1.

When the system size is only a few spins (panels (a)-(d) of Fig.~\ref{fig:Husimi_comparison_N_5}), the coherent states covers a wider section of phase space, and hence, 
after the dynamics, they have a non-negligible overlap. However, as the system size is increased (panels (e)-(h) of Fig.~\ref{fig:Husimi_comparison_N_5}), the distribution of the state in 
phase space becomes narrower and the choice of initial state matters: Only those initial states in the chaotic region of the semiclassical limit 
experience a behavior that smears their distribution across a large section of phase space (panel (h) in Fig.~\ref{fig:Husimi_comparison_N_5}). 

In our study of scrambling in the quantum kicked-top, we use the same parameters as in the semiclassical case and explore different 
diagnostics of scrambling. Below, we describe them in more detail.

\section{Measures of quantum chaos}\label{sec:measures}
\subsection{Out-of-time-ordered correlators and quantum chaos}\label{sec:otoc-and-qc}

Classically, we understand chaos as the exponential sensitivity to perturbations to initial conditions. We can express such a 
sensitivity with the aid of a Poisson bracket as
\begin{align}\label{eqn:pb}
\{x(t),p(0)\}_{PB}\sim\mathrm{e}^{\lambda t}.
\end{align}
In quantum mechanics, we have observables that do not commute, and the dynamics of closed systems are unitary. Therefore, we extend the notion of 
sensitivity to initial perturbations as follows. Consider a system with initial state $\rho$ and evolving unitarily by the 
dynamics generated by a Hamiltonian $H$. Moreover, in order to extend Eq.~\eqref{eqn:pb} into the quantum case, we also use two 
initially commuting operators $W$ and $V$. With these ingredients, we define the following
\begin{align}
C(t):= \langle\left[W(t),V(0)\right]^\dagger \left[W(t),V(0)\right].
\end{align}
In principle, for a system with a chaotic semiclassical limit we expect $C(t)\rangle\sim\mathrm{e}^{2\lambda t}$.
Here, $W(t) = U^\dagger W U$, and $U=\exp(\mathrm{i} H t)$.
At $t=0$, $C(t)$ will be zero since the operators $W$ and $V$ initially commute. As the operators cease to commute because of the 
dynamics, the value of $C(t)$ will increase. If the dynamics and the initial state have no chaotic equivalent in the semiclassical 
limit, then the recurrence time for $C(t)$ will be comparatively small, and it will show revivals. However, for an appropriate initial 
state and Hamiltonian, there will be a persistent growth of $C(t)$ until it reaches a maximum value around which it will fluctuate. 

We can also understand the dynamics by using the \emph{out-of-time-ordered correlator}. It is given by
\begin{align}
F(t) &:= \left\langle W^\dagger(t)V^\dagger W(t) V \right\rangle \\
&= \mathrm{Tr}\left( W^\dagger(t) V^\dagger W(t) V \rho\right). \nonumber
\end{align}
In the special case in which $W$ and $V$ are unitary operators, it is straightforward to verify that
\begin{align}\label{eqn:comm_rel_otoc}
C(t) = 2\left(1-\mathrm{Re}(F(t))\right).
\end{align}
Using a similar analysis to the one we did with $C(t)$, we can see that qualitatively, it is the persistent smallness of $F(t)$ that 
indicates we are observing chaotic behavior. 

In the case of the kicked-top, we use the operators
\begin{align}
\label{eqn:VandW}
    V = W(0) = \exp\left(\frac{1}{\sqrt{2j}}J_y\right).
\end{align}

With this choice, and initial coherent states as given by \eqref{eqn:cohenrent_state_def}, we would expect that for initial states in the chaotic region
and a sufficiently large system size, the OTOC would exhibit persistent smallness but otherwise exhibit quasiperiodic revivals.
We will however further explore this behavior with the aid of quasiprobabilities related to OTOCs.

\subsection{The nonclassicality behind OTOC's quasiprobabilities}\label{sec:otoc-qp-nc}
In order to help identify relevant features of the dynamics of the kicked-top it is useful to expand the operators $W(t)$ and $V$ in terms of the projectors onto their eigenspaces. When doing so, we can rewrite the OTOC as an expectation value of the product of eigenvalues, $v_i, w_j$, and obtain
\begin{align}
F(t) = \sum_{v_1,w_2,v_2,w_3} v_1 w_2 v_2^* w_3^* \tilde{p}_t \left(v_1, w_2, v_2, w_3 \right) .
\end{align}
In the above, 
\begin{align}
\tilde{p}_t \left(v_1, w_2, v_2, w_3 \right) := 
\mathrm{Tr}\left( \Pi_{w_3}^{W(t)} \Pi_{v_2}^{V} \Pi_{w_2}^{W(t)} \Pi_{v_1}^{V} \rho \right).
\end{align}
We call this quantity the (coarse-grained) quasiprobability behind the OTOC. From previous work \cite{GonzalezAlonso_OutofTimeOrderedCorrelatorQuasiprobabilitiesRobustly_2019} we define
the \emph{cumulative nonclassicality} of the OTOC quasiprobability distribution,
\begin{align}\label{eqn:nonclass_quasiprob}
\tilde{N}(t) := \sum_{v_1,w_2,v_2,w_3} \left| \tilde{p}_t \left(v_1, w_2, v_2, w_3 \right)  \right| - 1,
\end{align}
useful for detecting aspects of the dynamics that cannot be explained classically.

For the kicked-top model, we expect that initial states in the chaotic region will exhibit more nonclassicality after the unitary evolution generated by the Hamiltonian since we expect such states to be highly entangled.

\subsection{Tripartite Mutual Information}\label{sec:TMI}
As done in Ref.~\cite{Hosur_ChaosQuantumChannels_2016}, it is possible to consider unitary channels acting on $N$ qubit systems as 
states in a doubled Hilbert space and use the tripartite mutual information (TMI) involving partitions of the input and output spaces
as a measure of scrambling. If a unitary $U(t)$ is defined in the product basis by
\begin{align}\label{eqn:unitary-def}
U(t) = \sum_{m=1}^{2^N}\sum_{m'=1}^{2^N} u_{m,m'} \ket{m}\bra{m'}
\end{align}
then it is possible to also represent this as a $2N$-qubit state by taking the tensor product of both the input and output spaces of the 
channel
\begin{align}\label{eqn:unitary-as-state}
\ket{U(t)} = \frac{1}{2^{N/2}}\sum_{m=1}^{2^N}\sum_{m'=1}^{2^N} u_{m',m} \ket{m}_{in}\otimes\ket{m'}_{out}.
\end{align}
In particular, the identity channel can be written as
\begin{align}\label{eqn:identity-as-state}
\ket{\mathbb{I}} = \frac{1}{2^{N/2}}\sum_{m=1}^{2^N} \ket{m}_{in}\otimes\ket{m}_{out}.
\end{align}
Hence, it is possible to rewrite Eq.~\eqref{eqn:unitary-as-state} as
\begin{align}\label{eqn:unitary-as-state-compact}
\ket{U(t)} = \mathbb{I}\otimes U(t) \ket{\mathbb{I}}.
\end{align}
Using the mapping between unitaries and states, we can then partition the input into subsystems $A$ and $B$ and the output into 
subsystems $C$ and $D$ and define the tripartite mutual information for a channel as
\begin{align}\label{eqn:TMI-def}
I_3 (A:C:D) := I(A:C) + I(A:D) - I(A:CD).
\end{align}
The bipartite information between two subsystems, e.g. $I(A:C)$, is given by $I(A:C) = S_A + S_C - S_{AC}$. Here, each of the entropies 
requires the computation of a particular reduced density matrix. For instance, $S_{AC} = -\text{Tr}(\rho_{AC}\log_2 \rho_{AC})$ requires the
calculation of $\rho_{AC} = \mathrm{Tr}_{BD}\rho$. Here, $\rho$ is the density matrix $\ket{U(t)}\bra{U(t)}$ associated with the channel.
Using the entanglement entropies we can rewrite Eq.~\eqref{eqn:TMI-def} as
\begin{align}
\begin{split}
I_3 (A:C:D) &={} S_A + S_C + S_D \\&- S_{AC} - S_{AD} - S_{CD} + S_{ACD}.
\end{split}
\end{align}

While it is more common to calculate the TMI of a state of the system after the unitary evolution (as done for several many-body 
states in Refs.~\cite{Seshadri_TripartiteMutualInformation_2018,Iyoda_ScramblingQuantumInformation_2018}), the main motivation behind using 
the TMI of the state equivalent to the unitary evolution is that, channels that scramble will convert input states into locally 
indistinguishable output states. That is to say, the negativity of the tripartite mutual information in Eq.~\eqref{eqn:TMI-def} signals 
multipartite entanglement between the input $A$ and the output spaces $C$ and $D$, and hence, channels that scramble will have a 
persistent negativity close to maximal.

In our calculations of the TMI for the kicked-top model, we will use the partition: $A$ as one input spin, $B$ $N-1$ as remaining input spins, $C$ as one 
output spin, and $D$ as $N-1$ remaining output spins. Interestingly, with this partition our tripartite mutual information calculations 
simplify since $S_A = 1$, $S_C=1$, $S_D=1$, $S_{CD} = N$, and $S_{ACD} = N-1$. Therefore, only $S_{AC}$ and $S_{AD}$ will have 
nontrivial contributions to $I_3$ of the kicked-top unitary channel. 

\section{Numerical simulation results and discussion}\label{sec:numerics-results}

Here we calculate different measures of scrambling and nonclassicality as a funciton of time for initial states corresponding to the different initial points of the semiclassical phase space of Fig.~\ref{fig:stroboscopic_map_nk_e_50}. 

\begin{figure*}[!htbp]
    \subfloat[][Real part of the OTOC\label{fig:Re_OTOC_N_eq_5}]{
    \includegraphics[width=1.0\columnwidth]{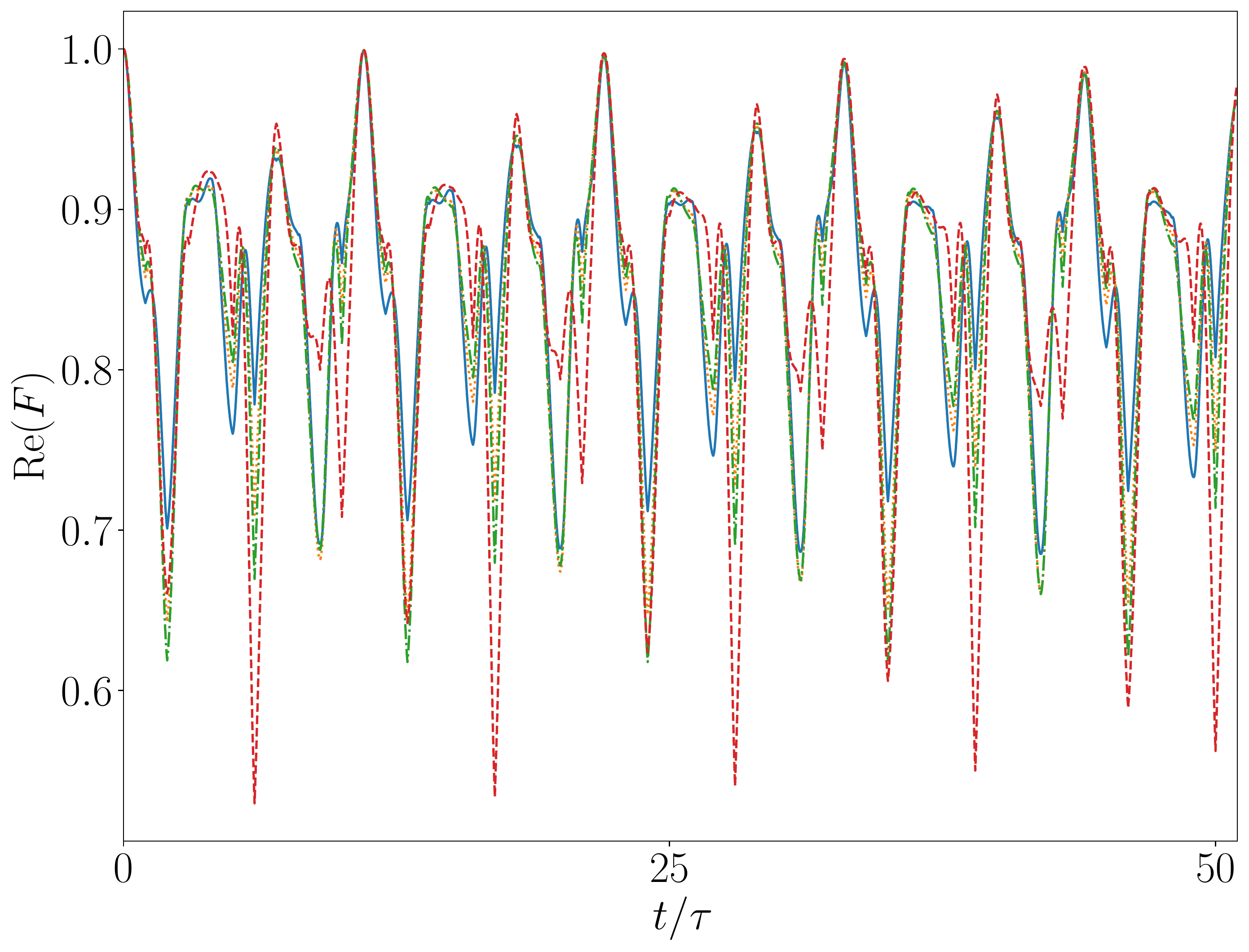}
    }
    \subfloat[Power spectrum of the real part of the OTOC \label{fig:FFT_Re_OTOC_N_eq_5}]{
    \includegraphics[width=0.985\columnwidth]{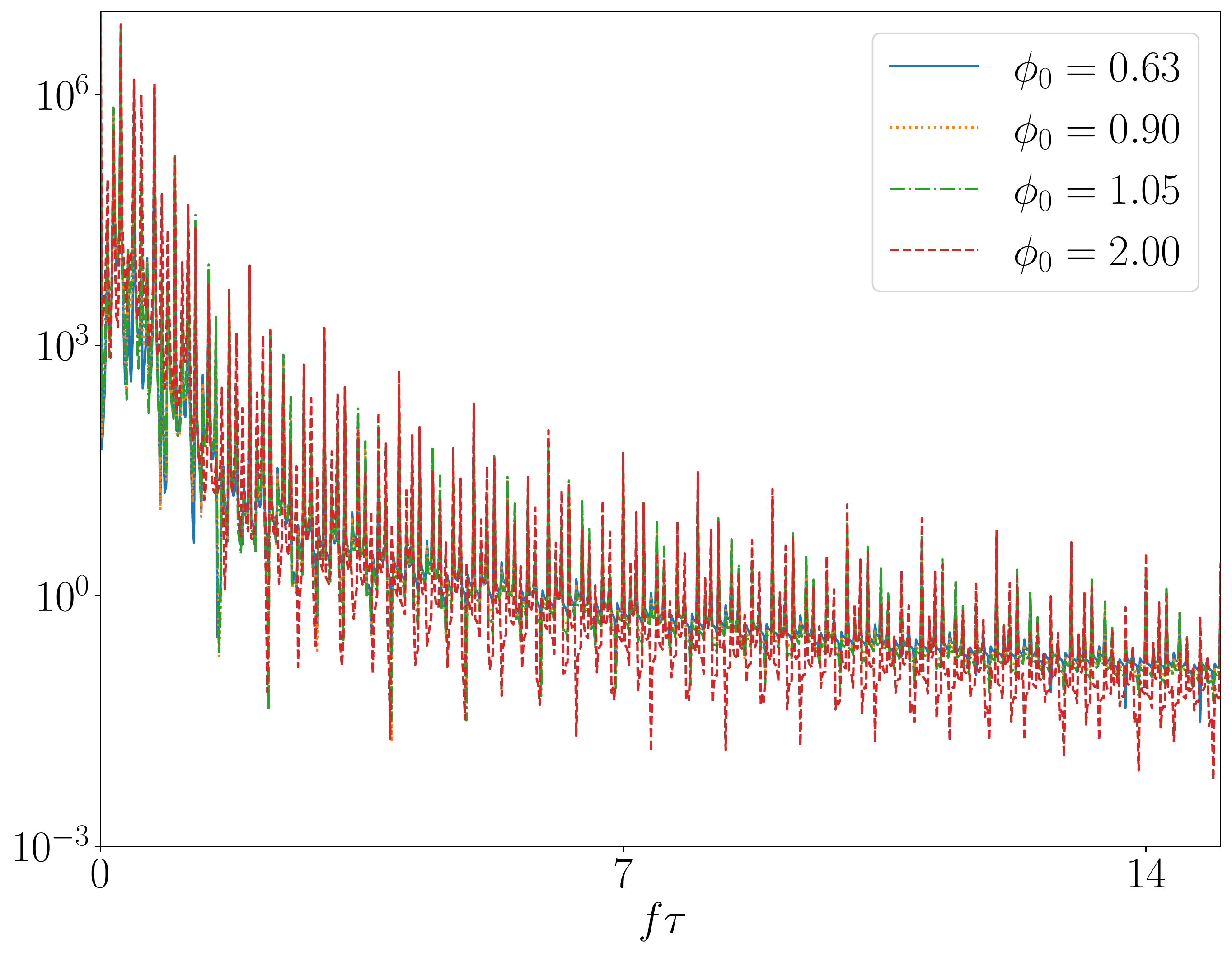}
    }
    \caption{(Color online) $\mathrm{Re}(F(t))$ and its power spectrum for a quantum kicked-top with $N=5$ spins and $\kappa=3.0$, $p=\pi/2$, and $\tau=1.0$ for several initial conditions and $n_k=50$ kicks. We use the following initial coherent states $\ket{\theta_0,\phi_0}$:
    (a) \emph{elliptic fixed point} with $\theta_0 = 2.25$, $\phi_0 = 0.63$, (b) \emph{regular region point} with 
    $\theta_0 = 2.25$, $\phi_0 = 0.90$, (c) \emph{edge of chaos point} with $\theta_0 = 2.25$,
    $\phi_0 = 1.05$, and (d) \emph{sea of chaos point} with $\theta_0 = 2.25$, $\phi_0 = 2.0$.}
    \label{fig:OTOC_and_FFT_comparison_N_5}
\end{figure*}

Taking advantage of the permutational symmetry of the problem, we make use of the numerical 
libraries in \cite{Shammah_OpenQuantumSystems_2018} as implemented in 
\cite{Johansson_QuTiPOpensourcePython_2012,Johansson_QuTiPPythonFramework_2013}
to calculate the OTOC and the nonclassicality of its quasi-probability distribution using the Dicke basis. The benefit of this approach is that it is possible to obtain results 
even for larger system sizes (like $N=100$) that would otherwise be too challenging to compute. We explain the details of the simulation in 
appendix \ref{sec:sim-details}, and present our results in Figs.~\ref{fig:OTOC_and_FFT_comparison_N_5} and \ref{fig:TMI_and_entropies}.

\begin{figure*}[!htbp]
    \subfloat[][Real part of the OTOC\label{fig:Re_OTOC_N_eq_100}]{
    \includegraphics[width=1.0\columnwidth]{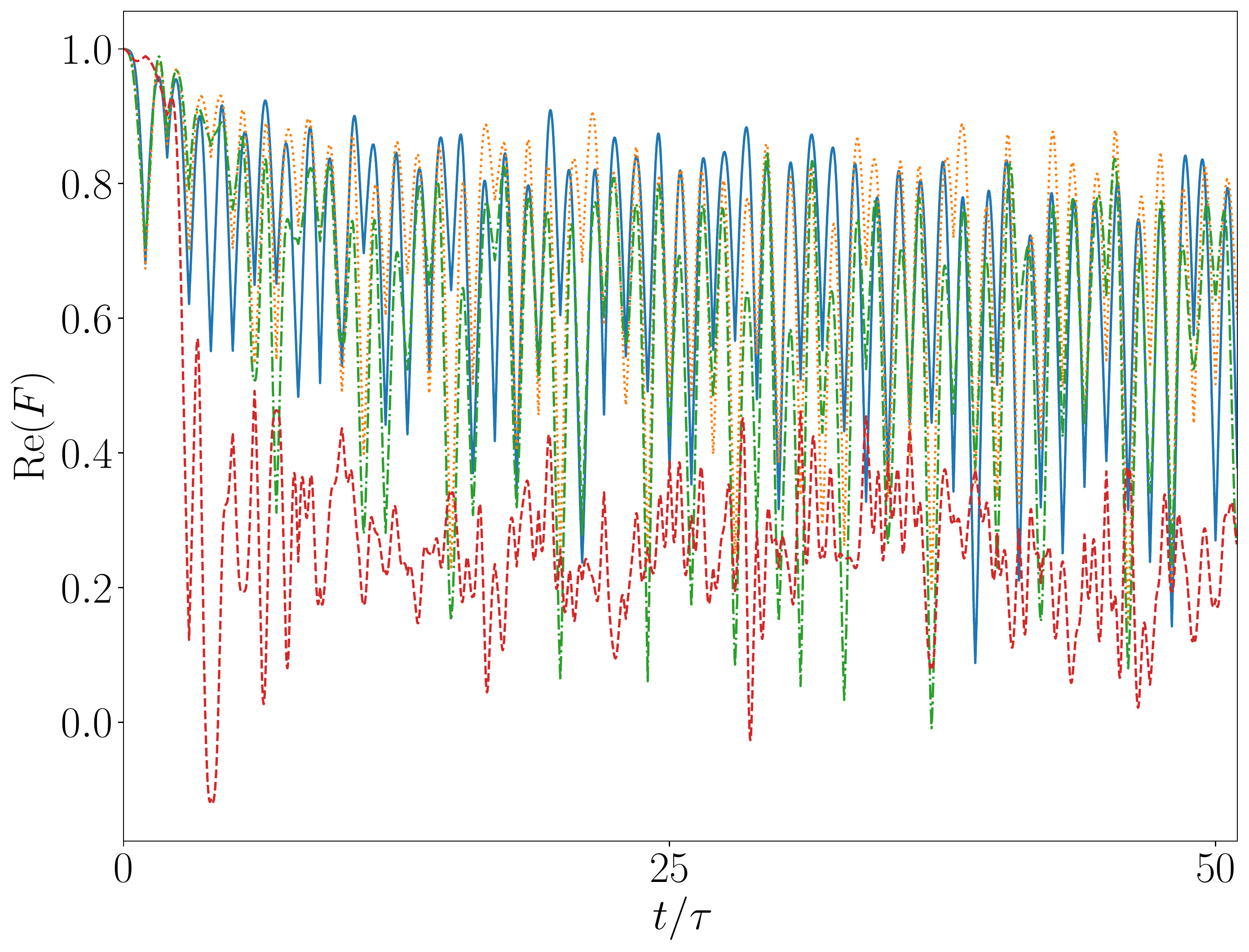}
    }
    \subfloat[Power spectrum of the real part of the OTOC \label{fig:FFT_Re_OTOC_N_eq_100}]{
    \includegraphics[width=.967\columnwidth]{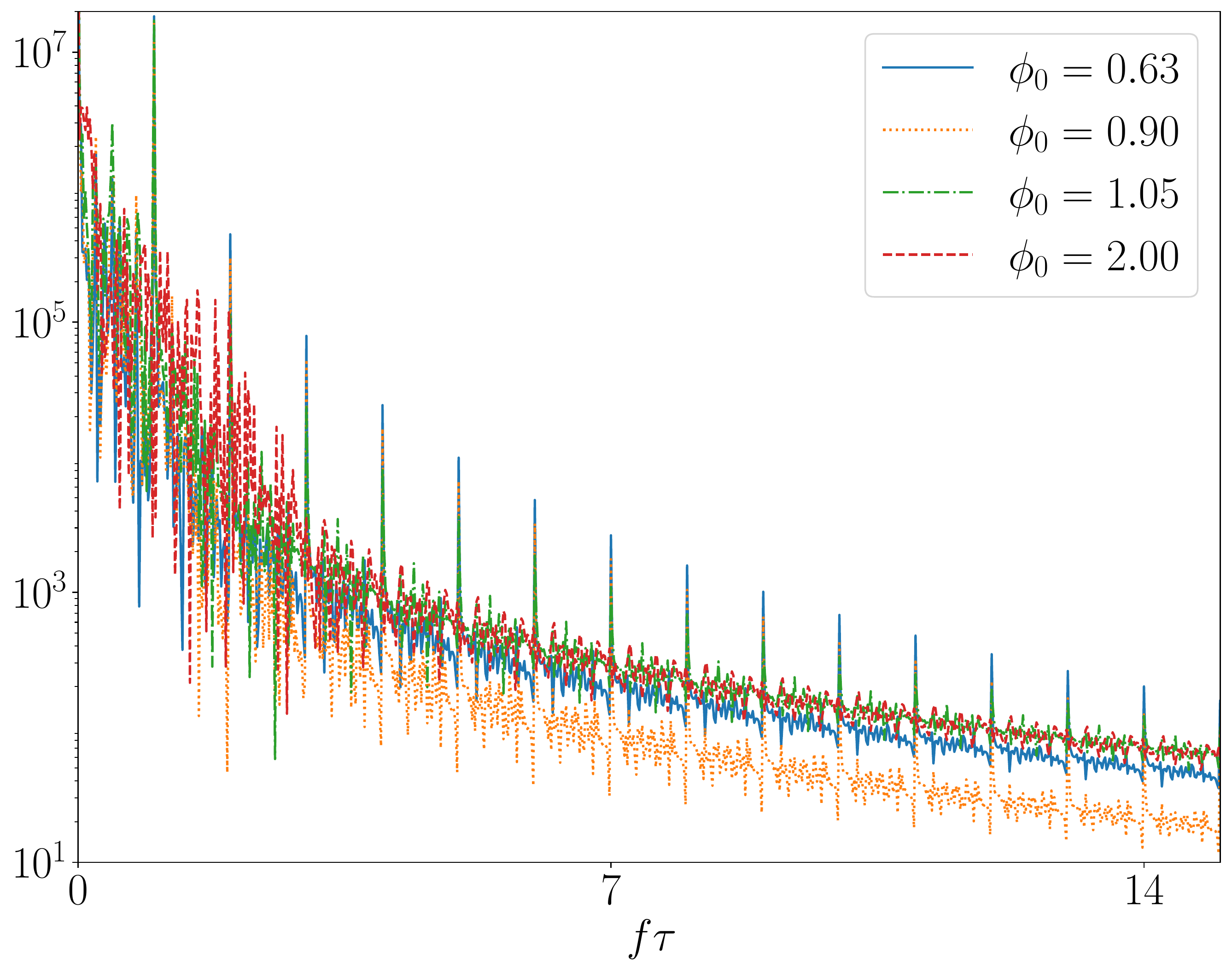}
    }
    \caption{(Color online) $\mathrm{Re}(F(t))$ and its power spectrum for a quantum kicked-top with $N=100$ spins and $\kappa=3.0$, $p=\pi/2$, and $\tau=1.0$ for several initial conditions and $n_k=50$ kicks. We use the following initial coherent states $\ket{\theta_0,\phi_0}$:
    (a) $\theta_0 = 2.25$, $\phi_0 = 0.63$, (b) $\theta_0 = 2.25$, $\phi_0 = 0.90$, (c) $\theta_0 = 2.25$,
    $\phi_0 = 1.05$, and (d) $\theta_0 = 2.25$, $\phi_0 = 2.0$.}
    \label{fig:OTOC_and_FFT_comparison_N_100}
\end{figure*}

In Figs.~\ref{fig:OTOC_and_FFT_comparison_N_5} to \ref{fig:OTOC_and_FFT_comparison_N_100} we can appreciate the real part of the OTOC 
and its power spectra for two different system sizes: $N=5$ and $N=100$. The power spectra of panel (b) are calculated simply with a discrete Fourier transform from the time-domain numerical data of panel (a). As is clear from the figures, in the case of the smaller system 
size, regardless of the initial state, it is not possible to observe a substantial difference between the different regions in phase 
space: both the real part of the OTOC and its power spectrum show no initial state dependence. As the system size is increased, the difference between the choice of 
initial state becomes more evident: 
The real part of the OTOC slowly decreases for all initial points. The peaks in the nonclassicality, corresponding to the kicks in the Hamiltonian, keep a hierarchy of the phase space points relation to chaos: the elliptic fixed point [$\phi/\pi=0.63$ (blue curves)] is greater than the regular region  [$\phi/\pi=0.90$ (orange curves)], which is greater than the point at the edge of the sea of chaos [$\phi/\pi=1.05$ (green curves)], which is greater than the point in the sea of chaos [$\phi/\pi=2.0$ (red curves)]. 

\begin{figure*}[!htbp]
    \subfloat[][Nonclassicality of the quasiprobability distribution behind the OTOC\label{fig:NC_QP_OTOC_N_eq_5}]{
    \includegraphics[width=1.0\columnwidth]{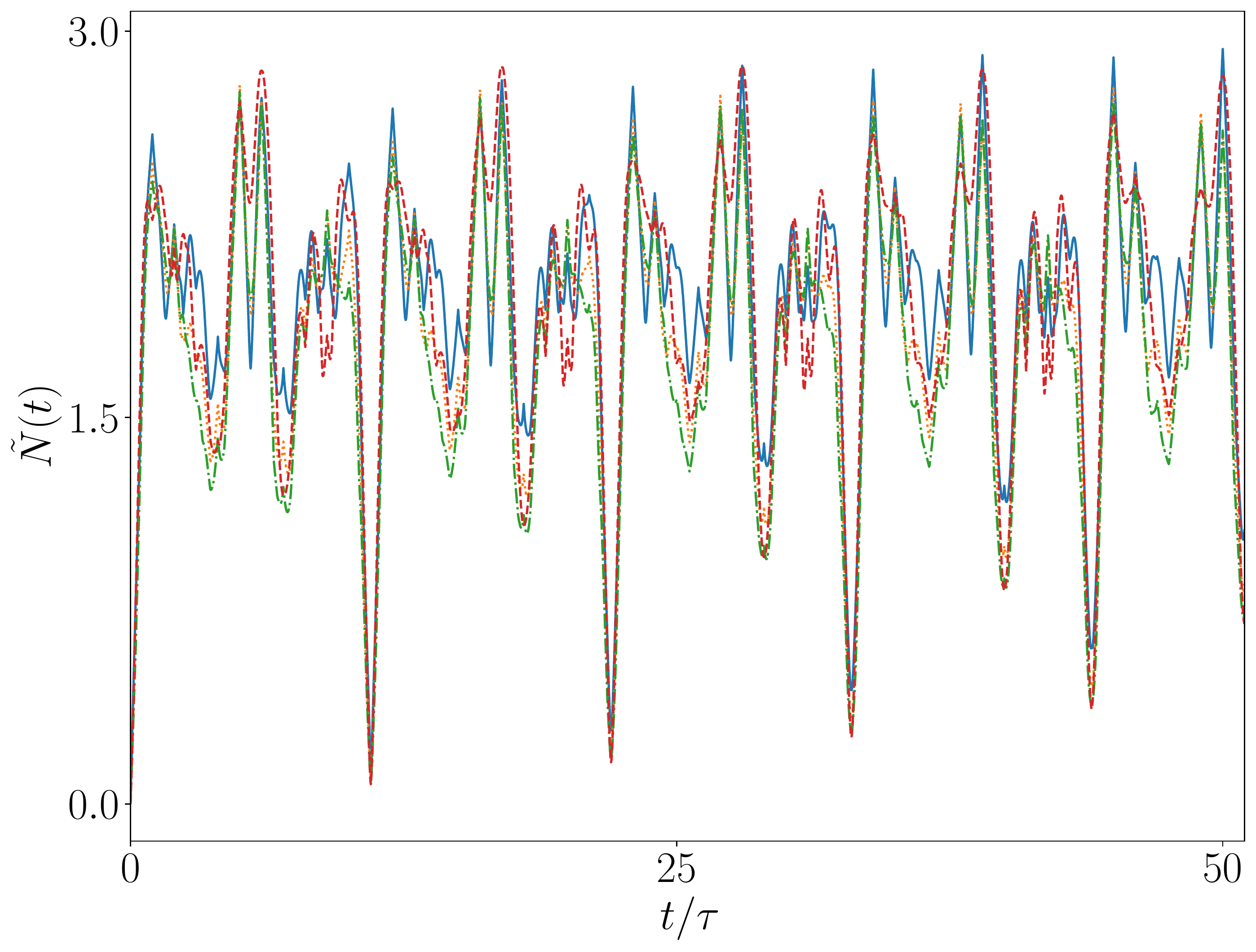}
    }
    \subfloat[Power spectrum of the nonclassicality of the quasiprobability distribution behind the OTOC \label{fig:FFT_NC_QP_OTOC_N_eq_5}]{
    \includegraphics[width=1.0\columnwidth]{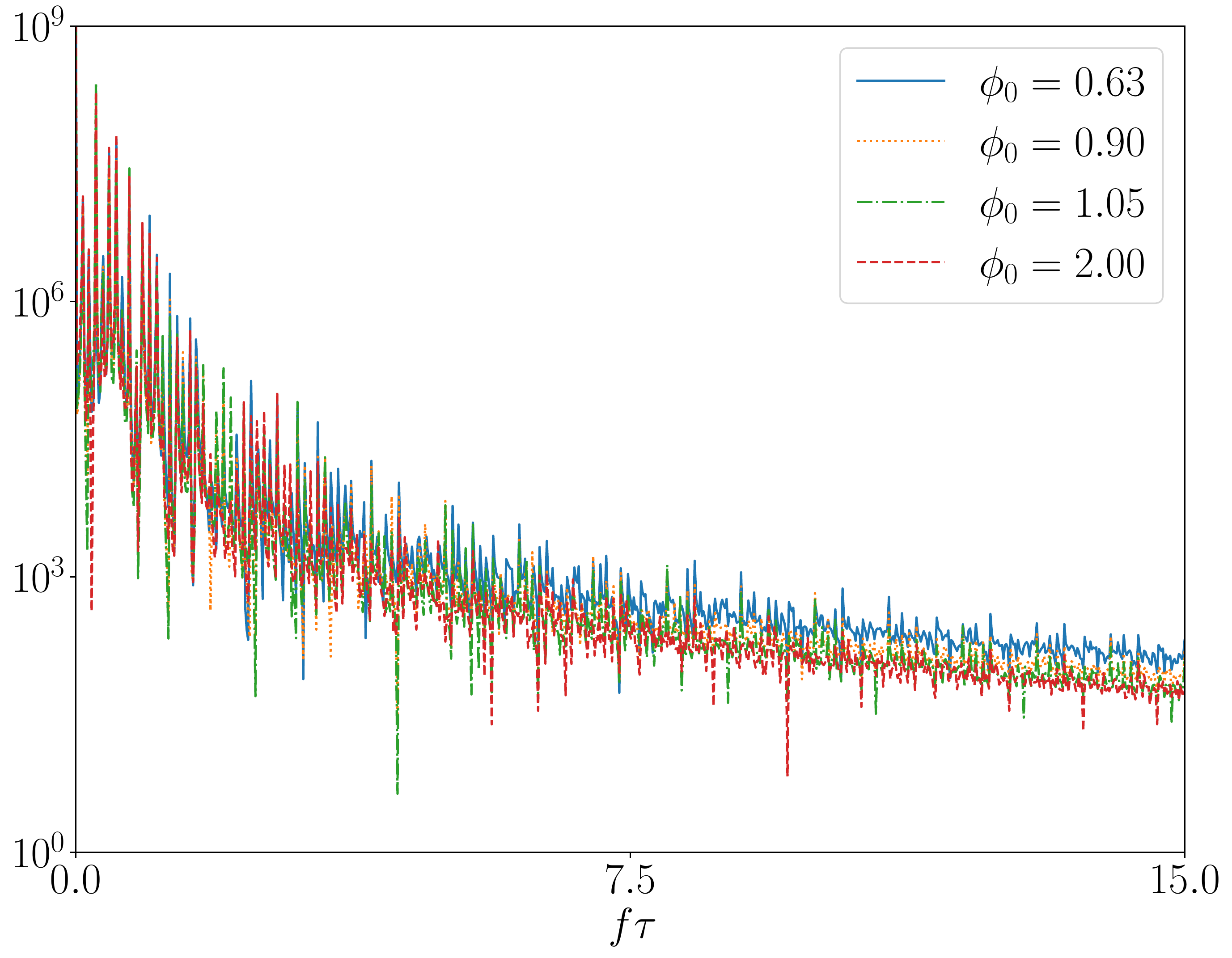}
    }
    \caption{(Color online) Nonclassicality of the quasiprobability behind the OTOC and its power spectrum for a quantum kicked-top with $N=5$ spins and $\kappa=3.0$, $p=\pi/2$, and $\tau=1.0$ for several initial conditions and $n_k=50$ kicks. We use the following initial coherent states $\ket{\theta_0,\phi_0}$:
    (a) $\theta_0 = 2.25$, $\phi_0 = 0.63$, (b) $\theta_0 = 2.25$, $\phi_0 = 0.90$, (c) $\theta_0 = 2.25$,
    $\phi_0 = 1.05$, and (d) $\theta_0 = 2.25$, $\phi_0 = 2.0$.}
    \label{fig:NC_QP_OTOC_and_FFT_comparison_N_5}
\end{figure*}

\begin{figure*}[!htbp]
    \subfloat[][Nonclassicality of the quasiprobability distribution behind the OTOC\label{fig:NC_QP_OTOC_N_eq_100}]{
    \includegraphics[width=1.0\columnwidth]{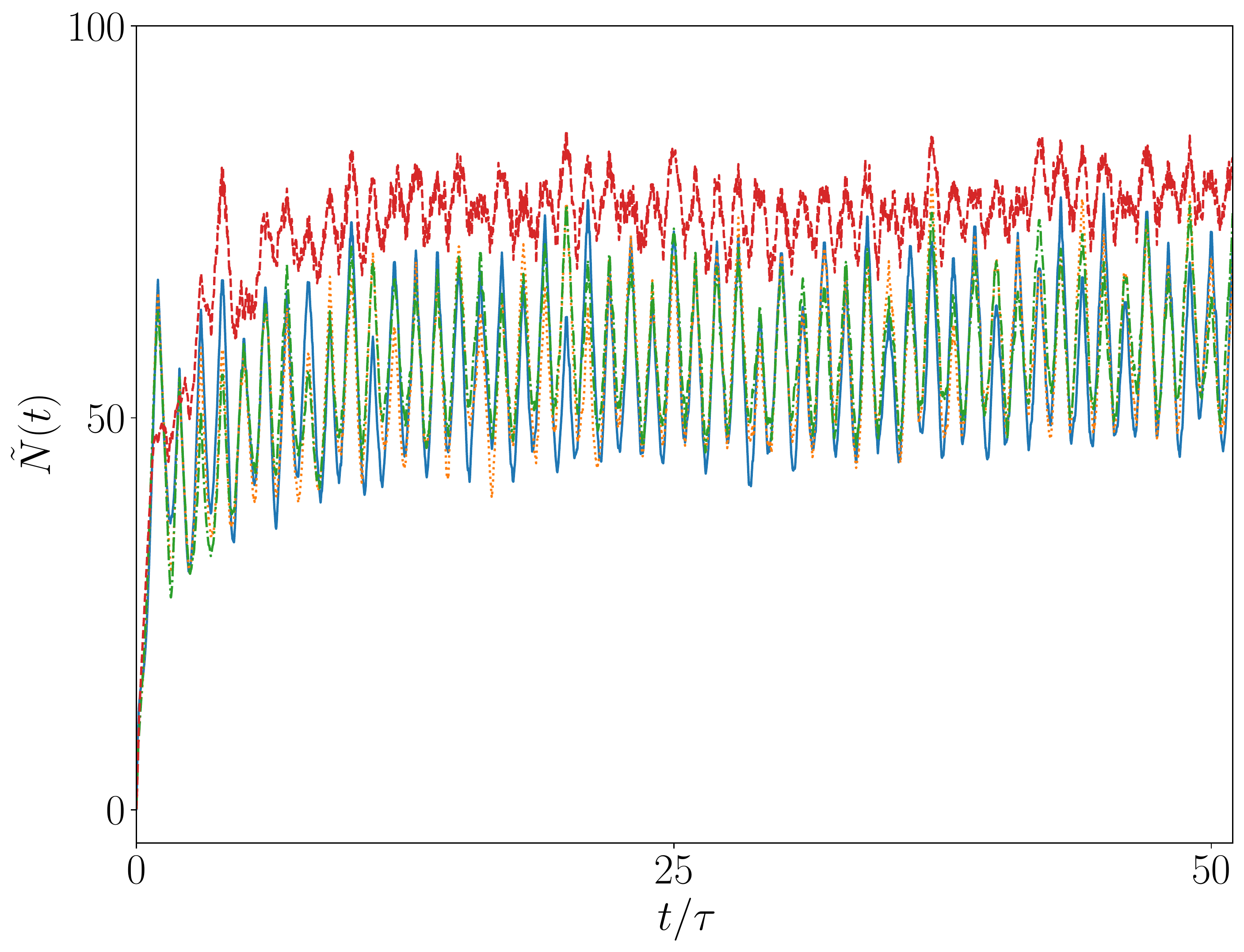}
    }
    \subfloat[Power spectrum of the nonclassicality of the quasiprobability distribution behind the OTOC \label{fig:FFT_NC_QP_OTOC_N_eq_100}]{
    \includegraphics[width=1.0\columnwidth]{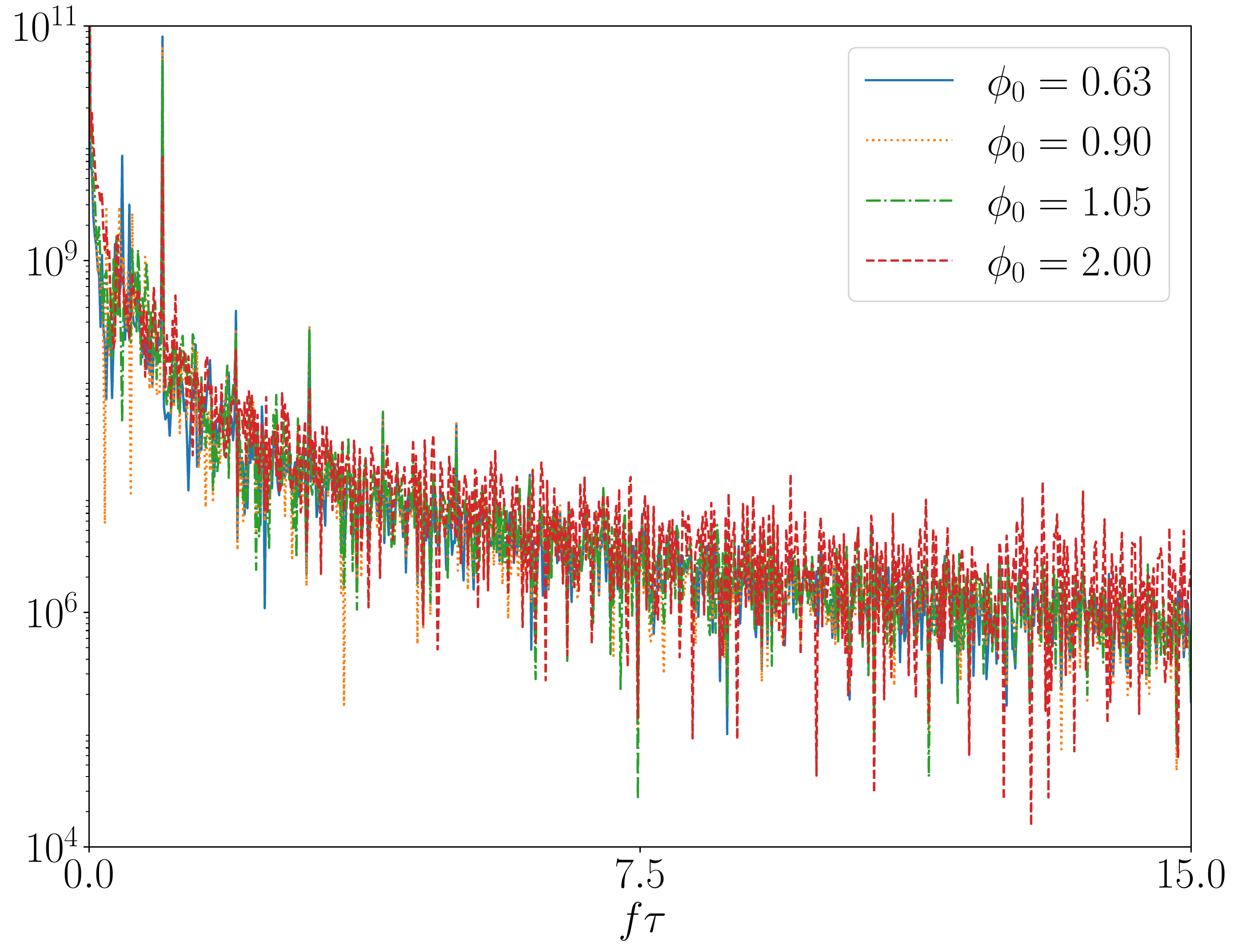}
    }
    \caption{(Color online) Nonclassicality and its power spectrum for a quantum kicked-top with $N=100$ spins and $\kappa=3.0$, $p=\pi/2$, and $\tau=1.0$ for several initial conditions and $n_k=50$ kicks. We use the following initial coherent states $\ket{\theta_0,\phi_0}$:
    (a) $\theta_0 = 2.25$, $\phi_0 = 0.63$, (b) $\theta_0 = 2.25$, $\phi_0 = 0.90$, (c) $\theta_0 = 2.25$,
    $\phi_0 = 1.05$, and (d) $\theta_0 = 2.25$, $\phi_0 = 2.0$.}
    \label{fig:NC_QP_OTOC_and_FFT_comparison_N_100}
\end{figure*}

In 
the nonclassicality in Figs.~\ref{fig:NC_QP_OTOC_and_FFT_comparison_N_5} and \ref{fig:NC_QP_OTOC_and_FFT_comparison_N_100} for small 
systems, the difference between the behavior of the various initial states is difficult to distinguish. However, for a large system size the nonclassicality 
for the initial chaotic points has a slow onset towards a value around which it will later oscillate. This value is larger than the corresponding ones for the other initial states. Additionally, we also observe the effect that the system size 
has on the power spectra of the OTOC and the nonclassicality of its quasi-probability distribution -- in all cases a $1/f$ noise type of behavior is observed.

For large system sizes and initial states in the chaotic sea, the kick harmonics are suppressed and there is an increased noise 
floor. This is to be expected of chaotic behavior that erases any signature of periodicity from the Hamiltonian. However, unlike the 
power spectra of the OTOC, the power spectra of the nonclassicality show little difference between the choice of initial state.

\begin{figure*}[!htbp]
    \subfloat[][$N=5$\label{fig:S_VN_N_eq_5}]{
    \includegraphics[width=.985\columnwidth]{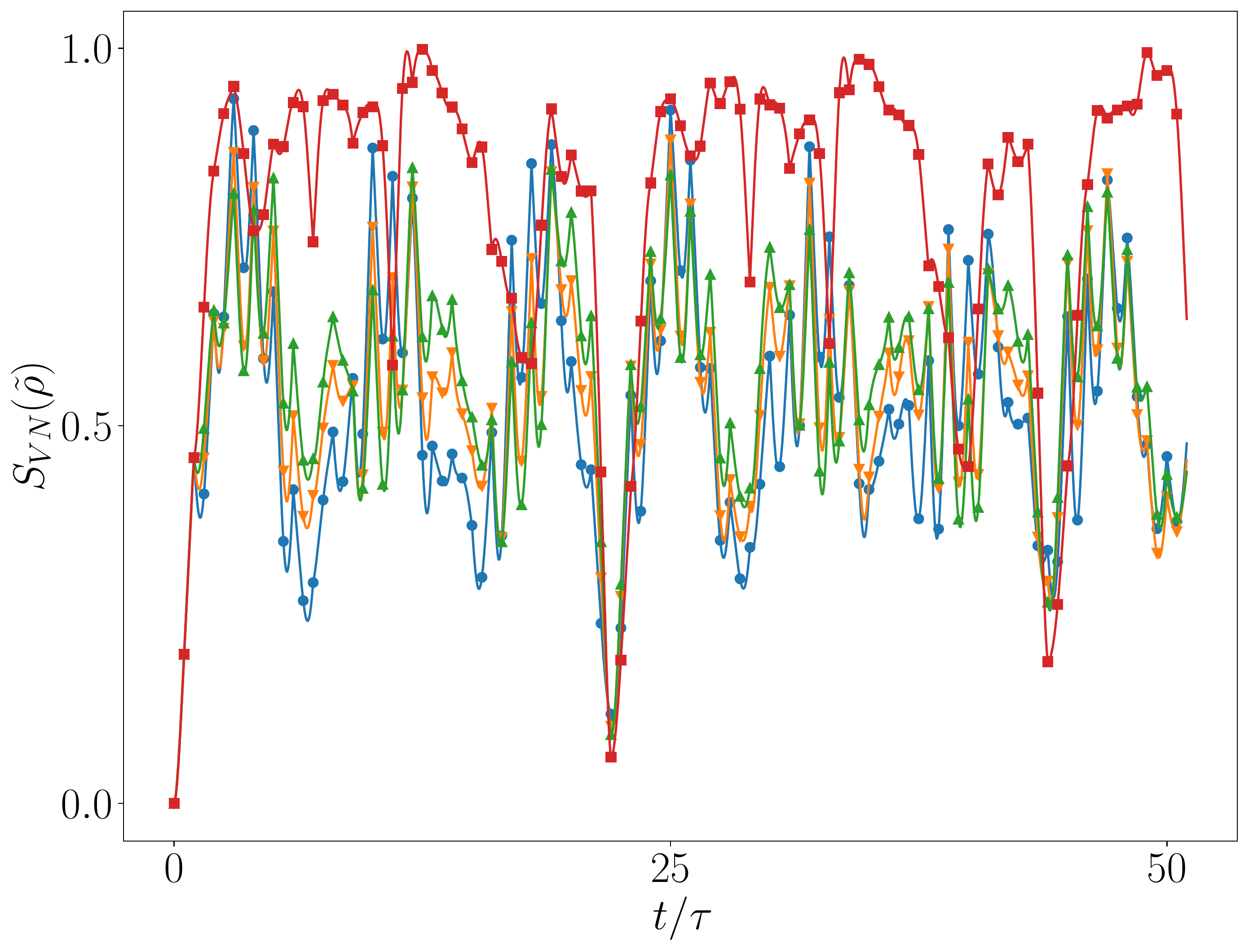}
    }
    \subfloat[][$N=100$\label{fig:S_VN_N_eq_100}]{
    \includegraphics[width=1.0\columnwidth]{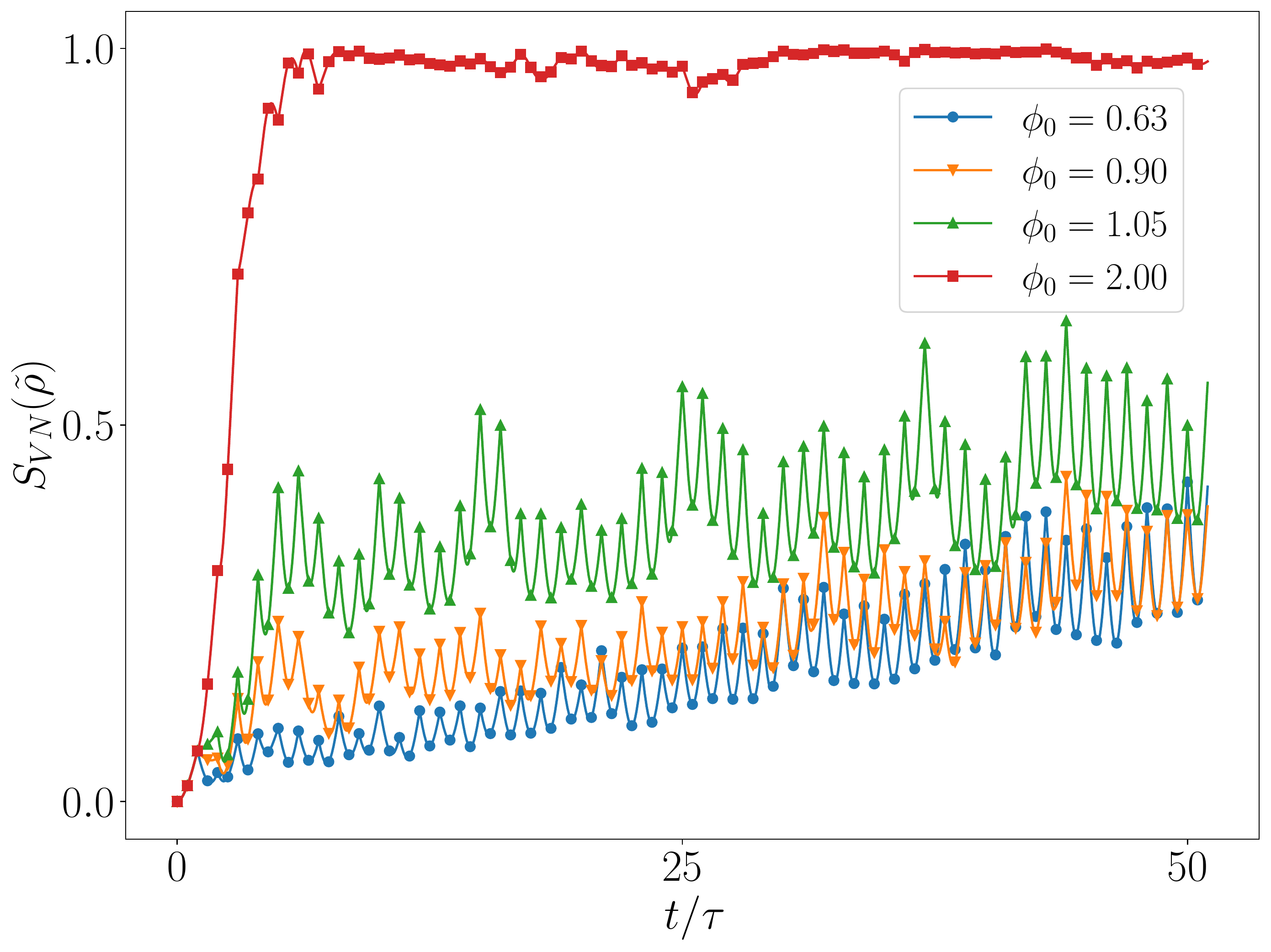}
    }
    \caption{(Color online) Von Neumann entanglement entropy for the reduced density matrix of one spin in the kicked-top after evolving for a time $t$.
    Shown is the behavior with $\kappa=3.0$, $p=\pi/2$, $\tau=1.0$. Different initial spin coherent states $\ket{\theta_0,\phi_0}$ have 
    the same value of $\theta_0 = 2.25$. By varying the value of $\phi_0$ different types of behavior can be studied going form a point 
    in the regular region of phase space $(\phi_0 = 0.63)$ to a point in the sea of chaos $(\phi_0 = 2.0)$. While the entanglement 
    entropy is largest for the initial state in the sea of chaos, it is only after making the system size large that the difference 
    between initial states becomes clear. Furthermore, in the chaotic case for a large $N$ the signatures of the periodic nature of the Hamiltonian are lessened and it remains close to its theoretical maximal value.}
    \label{fig:S_VN_comparison}
\end{figure*}

\begin{figure*}[!htbp]
    \includegraphics[width=1\columnwidth]{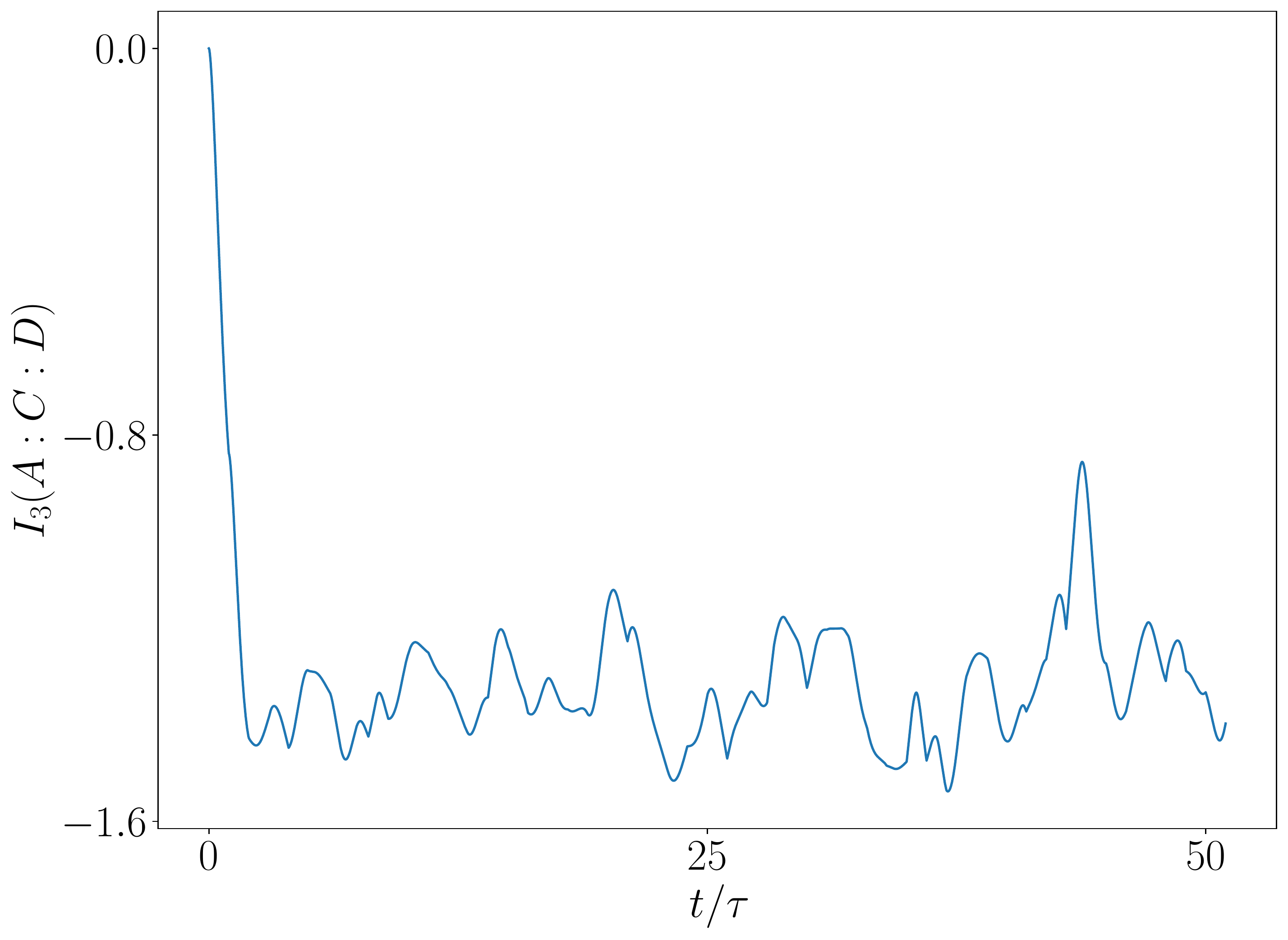}
    \includegraphics[width=0.95\columnwidth]{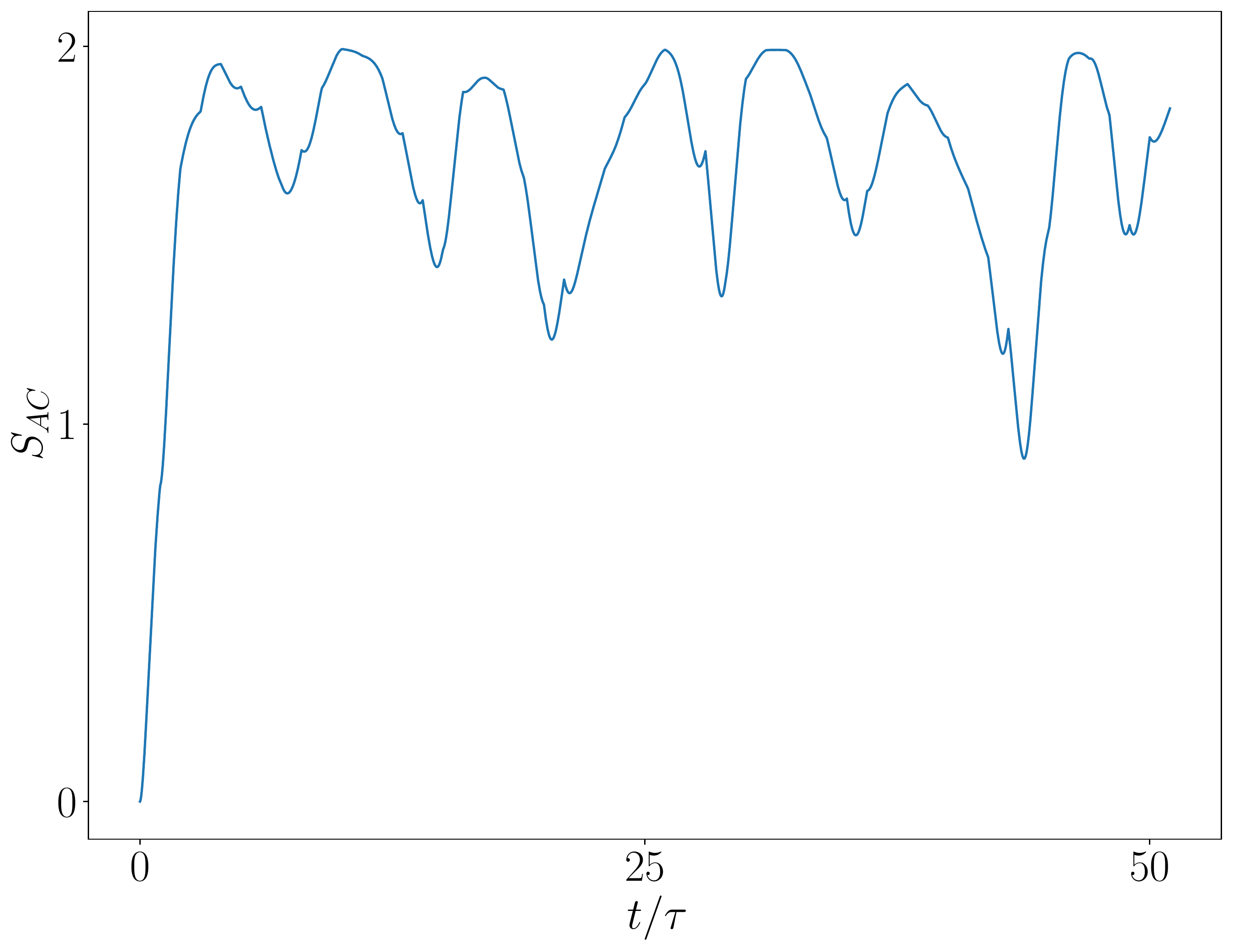}
    \includegraphics[width=0.95\columnwidth]{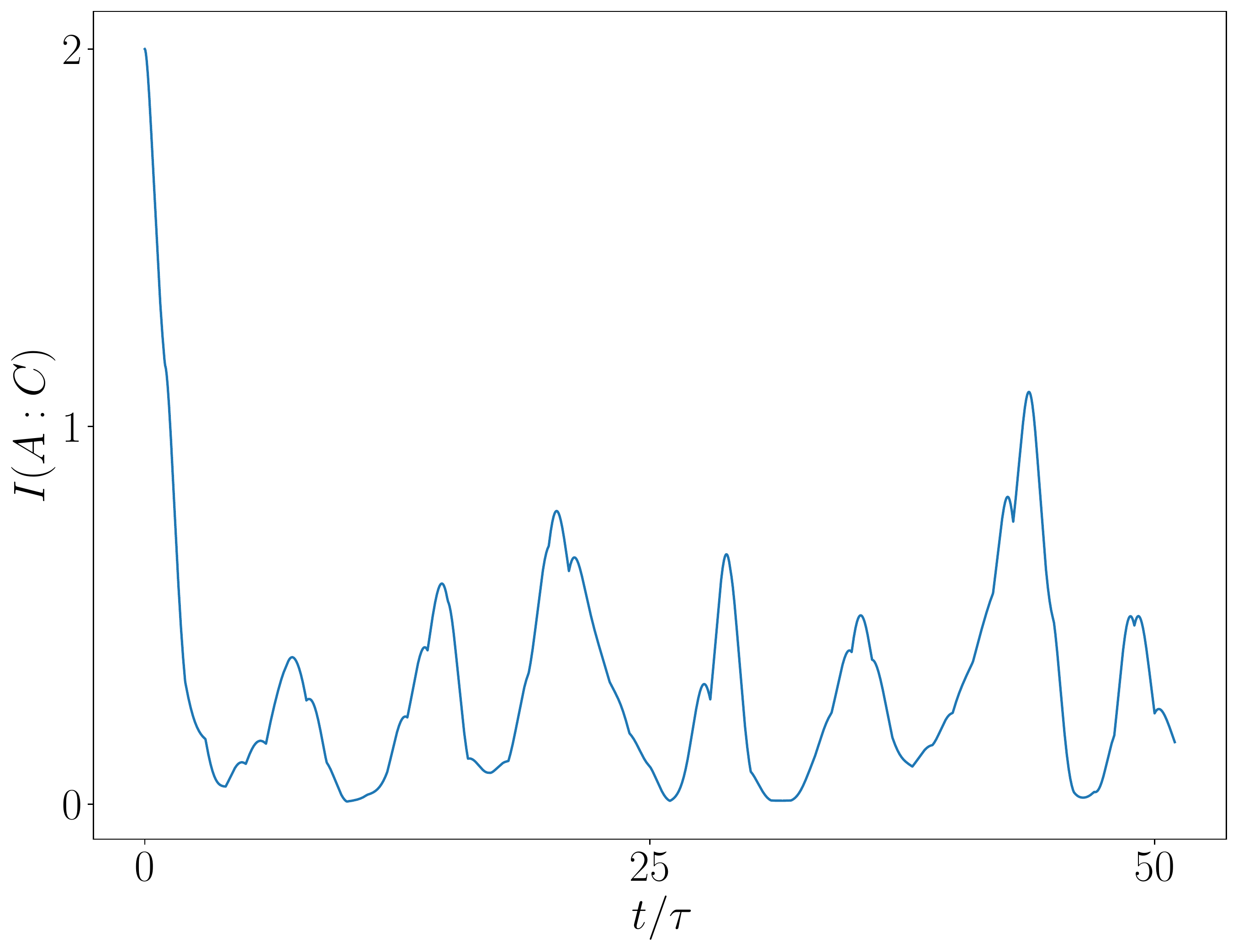}
    \includegraphics[width=0.97\columnwidth]{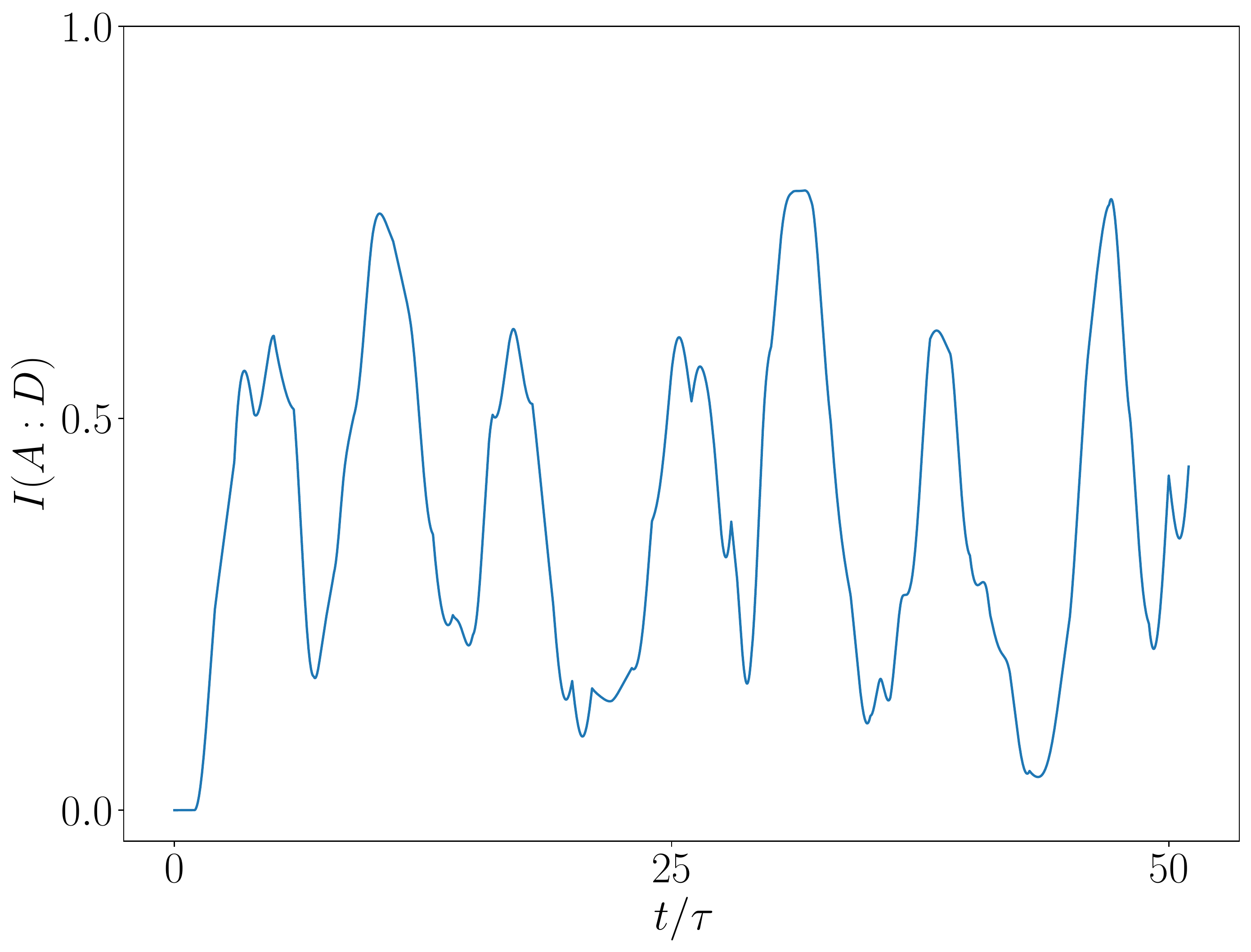}
    \caption{(Color online) Analysis of chaos and scrambling in the quantum kicked-top with $\kappa=3.0$, $p=\pi/2$, $\tau=1.0$, and $N=5$ for the $1:N-1$ 
    partition for both input and output of the unitary channel generated by the Hamiltonian. Given our choice of partition, $I_3$
    is a trivial function of $S_{AC}$ and $S_{AD}$. We appreciate in (a) that the tripartite 
    mutual information $I_3$ shows persistent negativity with fluctuations that depend on the size of the system, and in (b) that $S_{AC}$ oscillates quasi-periodically below 
    its maximum value. We see also a conservation of information from the plots of the mutual information between subsystems $A$ and $C$ 
    (in subplot(c) and $A$ and $D$ (in subplot (d)). Whenever $I(A:C)$ reaches a maximum, $I(A:D)$ is at a minimum and vice versa. The oscillations present in all plots are due to the periodic nature of the Hamiltonian and they disappear as $N$ becomes larger.}
    \label{fig:TMI_and_entropies}
\end{figure*}

We now contrast the behavior of the OTOC, the nonclassicality of its quasiprobability distribution, and their power spectra with that of the entropy of entanglement for the reduced density matrix of one spin. In the case of the kicked-top there is evidence \cite{Neill_ErgodicDynamicsThermalization_2016,Ruebeck_EntanglementItsRelationship_2017,Ghose_QuantumChaosTunneling_2008,Ghose_ChaosEntanglementDecoherence_2008,Kumari_UntanglingEntanglementChaos_2019,Kumari_QuantumclassicalCorrespondenceVicinity_2018} that 
indeed, the entanglement between one spin and the rest should have a similar behavior to what we have seen in the OTOC. We observe in 
Fig.~\ref{fig:S_VN_comparison} that as we make the system size larger the difference between the initial states becomes more obvious, 
with the entanglement between one spin and the rest for the system with $N=100$ spins being the largest when the initial state is in the 
sea of chaos. Furthermore, for an initial state in the chaotic region, we can also appreciate how by increasing the system size, the 
quasiperiodic oscillations in the entanglement entropy are also inhibited. 

Finally, the results we obtain for the TMI of the channel offer similar conclusions to those of the OTOC, the nonclassicality of its quasi-probability distribution, and the 
entanglement entropy of the reduced density matrix of one qubit regarding the effects of the size of the system.
In Fig.~\ref{fig:TMI_and_entropies} we can observe the TMI as well as the different entropies (and bipartite measures of mutual 
information) for the $N=5$ case. The persistent negativity of the TMI signals a scrambling channel. However, we can also appreciate the effects of a small system size in 
the large fluctuations of the $I_3 (A:C:D)$ and $S_{AC}$, $I(A:C)$, and $I(A:D)$. As the system grow larger, such fluctuations are suppressed and the TMI gets closer to its lowest negative value.

Typically, scrambling is studied in the context of systems with a notion of spatial locality. In such a case, under unitary time 
evolution generated by an appropriate Hamiltonian $H$, initially local operators evolve into sums of products involving many high weight
terms, that is, many operators acting on a high number of sites in the system. However, such an intuition is not adequate for systems,
such as the kicked-top, where the operators of interest act collectively. From our discussion of the kicked-top model in Sec.~\ref{sec:kicked-top}, we 
can see then that scrambling is understood using phase space as follows: Operators and states that have distributions in phase space that
are initially local are then smeared by the unitary chaotic evolution across a wide section of phase space with a long recurrence time. 
This explains both the persistent smallness of the OTOC and the TMI: as states and operators become less local in phase space then the square of their commutators with a fixed operator grows larger. 

Additionally, we also see how the size of the system affects the 
entanglement entropy of the reduced density matrix of one spin. As was previously observed in \cite{Kumari_UntanglingEntanglementChaos_2019}, the entanglement of the $1:N-1$ partition depends on both the dimensions of the space and how close the dynamics leave a state to a 
spin coherent state. This explains why we observe higher entanglement, and thus a higher nonclassicality, for states in the 
chaotic region - the dynamics takes them away from spin coherent states.

\section{Conclusions and outlook}\label{sec:conclusions}
We see that the notion of scrambling as measured by the OTOC, the nonclassicality of its quasi-probability distribution, the TMI of the channel, or the entanglement entropy
of the $1:N-1$ partition are all affected by the system size. When the system size is small, regardless of the measure used, the results can be misleading. In the case of the OTOC, the nonclassicality of its quasi-probability distribution, and the entanglement entropy we observe that for small system sizes the 
initial state makes little difference in the observed behavior. 

Additionally, we can appreciate that all of these measures still exhibit 
quasiperiodic behavior even for initial states in the sea of chaos. In the case of the TMI, we also observe fluctuations that are  
reminiscent of the periodic nature of the Hamiltonian. However, as the system size increases, the OTOC, the nonclassicality of its quasi-probability distribution, and the entanglement entropy all clearly show signatures of scrambling behavior only for initial states in the sea of chaos of the semiclassical limit of phase space.

Furthermore, for such states the fluctuations due to the periodic nature of the Hamiltonian are attenuated.
We understand this behavior as being related to the delocalization of states in phase space. As the system size increases, 
the volume of phase space occupied by the initial states is smaller, and the dynamics will only smear the state across a large section 
of the phase space when the initial state is in the sea of chaos. Conversely, when the system size is small, most initial states will cover 
a wide section of the phase space, and the dynamics spreads the states in a very similar fashion. Thus, we can have a rather intuitive 
understanding of scrambling dynamics even in systems that lack a notion of spatial locality: scrambling is the delocalization of states
in phase space.

Additionally, our comparison of the different tools to diagnose scrambling and quantum chaos highlights that while the channel TMI may 
start to signal scrambling at smaller system sizes than the OTOC, the nonclassicality of its quasi-probability distribution, or the entanglement entropy of the $1:N-1$ 
partition, it is also the most difficult to compute: it requires doubling the size of our space, and even in systems with a high degree
of symmetry, this makes its computation challenging.

On the other hand, while the entanglement entropy of the $1:N-1$ partition might be the easiest to calculate, it would be the most
affected by imperfections in an experimental setting. The calculation of the OTOC and the nonclassicality can be made efficiently by using the 
symmetries of the problem. Additionally, the nonclassicality is more robust than the OTOC, or the entanglement entropy, in the presence of 
experimental imperfections \cite{YungerHalpern_QuasiprobabilityOutofTimeOrderedCorrelator_2018}.

Our study highlights some future research opportunities. We have seen that the power spectrum of the OTOC and thenonclassicality of its quasi-probability distribution exhibit
fluctuations that are attenuated at larger system sizes and for states in the chaotic region of phase space. So far the behaviors of 
these fluctuations have been studied for initial thermal states \cite{Tsuji_OutoftimeorderFluctuationdissipationTheorem_2018,Tsuji_FluctuationTheoremQuantumstate_2018}. However, it should be possible to do so for more general initial states - the size of such 
fluctuations could be used to diagnose quantum chaos. Additionally, while we did not compute it, it would be interesting to study the quasi-probability distribution 
of the kicked-top model system and verify whether a branching behavior similar to the one observed in spin chains is also observed.

Our work complements previous work on the fluctuation-dissipation theorem \cite{Tsuji_OutoftimeorderFluctuationdissipationTheorem_2018,Tsuji_FluctuationTheoremQuantumstate_2018}, which illustrated a relation between chaotic dynamics in quantum systems and a nonlinear-response function in the OTOC. We further investigated the relation between the creation of entanglement and chaos in the quantum kicked-top, where in the comparison with the classical model provided by coherent spin states an upper bound for the Von Neumann entropy had been derived \cite{Kumari_UntanglingEntanglementChaos_2019,Kumari_QuantumclassicalCorrespondenceVicinity_2018}. 

Future work extending the current spin-system, finite-space analysis to the continuous-variable case through a phase-space analysis may investigate those systems' different scrambling domains in different modes \cite{Zhuang_ScramblingComplexityPhase_2019}. Additionally, another system that could be studied with the tools presented here is the family of kicked spin models in \cite{Munoz-Arias_NonlinearDynamicsQuantum_2021}.
An interesting opportunity would be to apply the current analysis to light-matter models, where long-range interactions can be mediated by a common field, such as in the Dicke model \cite{Lerose_OriginSlowGrowth_2018, Lerose20,Sinha2021_fingerprint}.
Another interesting extension would be to analyze the scrambling behavior of the kicked top using temporal quantum steering \cite{Lin_WitnessingQuantumScrambling_2020}. We leave such an analysis for future work.
\begin{acknowledgments}
JRGA wishes to thank Arjendu Pattanayak for helpful discussions.
JRGA was supported by a fellowship from the Grand Challenges Initiative at Chapman University.

JD was partially supported by the Army Research Office (ARO) grant No. W911NF-18-1-0178.
F.N. is supported in part by:
Nippon Telegraph and Telephone Corporation (NTT) Research,
the Japan Science and Technology Agency (JST) [via
the Quantum Leap Flagship Program (Q-LEAP),
the Moonshot R\&D Grant Number JPMJMS2061, and
the Centers of Research Excellence in Science and Technology (CREST) Grant No. JPMJCR1676],
the Japan Society for the Promotion of Science (JSPS)
[via the Grants-in-Aid for Scientific Research (KAKENHI) Grant No. JP20H00134 and the
JSPS-RFBR Grant No. JPJSBP120194828],
the Army Research Office (ARO) (Grant No. W911NF-18-1-0358),
the Asian Office of Aerospace Research and Development (AOARD) (via Grant No. FA2386-20-1-4069), and
the Foundational Questions Institute Fund (FQXi) via Grant No. FQXi-IAF19-06.
\end{acknowledgments}

\appendix

\section{Efficient Numerical Simulation of the OTOC's quasi-probability distribution nonclassicality}
\label{sec:sim-details}
The collective nonclassicality as defined by Eq.~\eqref{eqn:nonclass_quasiprob} might, at first, seem too complex a calculation to attempt. We need to compute, for every value of $t,$ $(N+1)^4$ different quasiprobability cases. However, 
it is possible to obtain $\tilde{N}$, in practice, in a time comparable to that of an OTOC calculation provided we make some simplifying assumptions. Since we are interested in using spin coherent states $\ket{\theta,\varphi}$ as the initial state of our system, then the cumulative nonclassicality becomes
\begin{equation}\label{eqn:redefine_nonclass}
\begin{aligned}
\tilde{N}(t)&=\sum_{v_1,w_2,v_2,w_3} \left| \braket{\theta,\varphi|\hat{U}_t^\dagger|w_3}\right| \left|\braket{w_3|\hat{U}_t|v_2}\right|\\&\times\left|\braket{v_2|\hat{U}_t^\dagger|w_2}\right| \left| \braket{w_2|\hat{U}_t|v_1}\right| \Big| \braket{v_1|\theta,\varphi}\Big| - 1
\end{aligned}
\end{equation}
We define the following
\begin{subequations}
\begin{align}
\vec{s} &:= \left[\braket{v|\theta,\varphi}\right] \\
\hat{M}_t &:= \left[\braket{w|\hat{U}_t|v}\right] \\
\vec{s}_t^T &:= \braket{\theta,\varphi|\hat{U}_t^\dagger|w} = \vec{s}^T \hat{M}_t^T
\end{align}
\end{subequations}
With these simplifications the cumulative nonclassicality of the quasiprobability behind the OTOC's quasi-probability distribution can be written as
\begin{align}
\tilde{N}(t)&= \left|\vec{s}_t^T\right| \left|\hat{M}_t\right| \left|\hat{M}_t^T\right| \left|\hat{M}_t\right| \left|\vec{s}\right| - 1
\end{align}
This suggests a straightforward procedure for calculating $\tilde{N}$
\begin{itemize}
    \item Define $dt$ as the length of the time step. 
    Preallocate $\vec{s}$, $\hat{A}=\begin{bmatrix}\bra{w}\\\vdots\end{bmatrix}$, and $\hat{B} = \left[\ket{v} \cdots\right]$
    \item Define $U_t=\mathbb{I}$
    \item For each time step $t$:
        \begin{itemize}
            \item If a kick needs to be applied update the accumulation of unitaries as follows
            \begin{align}
                U_t \leftarrow U_{\text{kick}} U_t
            \end{align}
            otherwise update with the twist term
            \begin{align}
                U_t \leftarrow U_{\text{twist}}(dt) U_t
            \end{align}
            \item Define $M_t = \hat{A}\hat{U}_t\hat{B}$.
            \item Compute $\vec{s}_t^T = \vec{s}^T \hat{M}_t^T$
        \end{itemize}
        \item Store the cumulative nonclassicality at time $t$
            \begin{align}
                \tilde{N}(t)&= \left|\vec{s}_t^T\right| \left|\hat{M}_t\right| \left|\hat{M}_t^T\right| \left|\hat{M}_t\right| \left|\vec{s}\right| - 1
            \end{align}
\end{itemize}
\vspace{1.8mm}
\bibliography{qkt_otoc_biblio}

\begin{thebibliography}{61}%
\makeatletter
\providecommand \@ifxundefined [1]{%
 \@ifx{#1\undefined}
}%
\providecommand \@ifnum [1]{%
 \ifnum #1\expandafter \@firstoftwo
 \else \expandafter \@secondoftwo
 \fi
}%
\providecommand \@ifx [1]{%
 \ifx #1\expandafter \@firstoftwo
 \else \expandafter \@secondoftwo
 \fi
}%
\providecommand \natexlab [1]{#1}%
\providecommand \enquote  [1]{``#1''}%
\providecommand \bibnamefont  [1]{#1}%
\providecommand \bibfnamefont [1]{#1}%
\providecommand \citenamefont [1]{#1}%
\providecommand \href@noop [0]{\@secondoftwo}%
\providecommand \href [0]{\begingroup \@sanitize@url \@href}%
\providecommand \@href[1]{\@@startlink{#1}\@@href}%
\providecommand \@@href[1]{\endgroup#1\@@endlink}%
\providecommand \@sanitize@url [0]{\catcode `\\12\catcode `\$12\catcode
  `\&12\catcode `\#12\catcode `\^12\catcode `\_12\catcode `\%12\relax}%
\providecommand \@@startlink[1]{}%
\providecommand \@@endlink[0]{}%
\providecommand \url  [0]{\begingroup\@sanitize@url \@url }%
\providecommand \@url [1]{\endgroup\@href {#1}{\urlprefix }}%
\providecommand \urlprefix  [0]{URL }%
\providecommand \Eprint [0]{\href }%
\providecommand \doibase [0]{http://dx.doi.org/}%
\providecommand \selectlanguage [0]{\@gobble}%
\providecommand \bibinfo  [0]{\@secondoftwo}%
\providecommand \bibfield  [0]{\@secondoftwo}%
\providecommand \translation [1]{[#1]}%
\providecommand \BibitemOpen [0]{}%
\providecommand \bibitemStop [0]{}%
\providecommand \bibitemNoStop [0]{.\EOS\space}%
\providecommand \EOS [0]{\spacefactor3000\relax}%
\providecommand \BibitemShut  [1]{\csname bibitem#1\endcsname}%
\let\auto@bib@innerbib\@empty
\bibitem [{\citenamefont {Larkin}\ and\ \citenamefont
  {Ovchinnikov}(1969)}]{Larkin_QuasiclassicalMethodTheory_1969}%
  \BibitemOpen
  \bibfield  {author} {\bibinfo {author} {\bibfnamefont {A.~I.}\ \bibnamefont
  {Larkin}}\ and\ \bibinfo {author} {\bibfnamefont {Y.~N.}\ \bibnamefont
  {Ovchinnikov}},\ }\bibfield  {title} {\enquote {\bibinfo {title}
  {Quasiclassical {{Method}} in the {{Theory}} of {{Superconductivity}}},}\
  }\href@noop {} {\bibfield  {journal} {\bibinfo  {journal} {Soviet Journal of
  Experimental and Theoretical Physics}\ }\textbf {\bibinfo {volume} {28}},\
  \bibinfo {pages} {1200} (\bibinfo {year} {1969})}\BibitemShut {NoStop}%
\bibitem [{\citenamefont
  {Kitaev}(2014)}]{Kitaev_HiddenCorrelationsHawking_2014}%
  \BibitemOpen
  \bibfield  {author} {\bibinfo {author} {\bibfnamefont {A.}~\bibnamefont
  {Kitaev}},\ }\href@noop {} {\enquote {\bibinfo {title} {Hidden correlations
  in the {{Hawking}} radiation and thermal noise},}\ } (\bibinfo {year}
  {2014}),\ \bibinfo {note} {talk given at {{Fundamental Physics Prize
  Symposium}}}\BibitemShut {NoStop}%
\bibitem [{\citenamefont {Shenker}\ and\ \citenamefont
  {Stanford}(2014{\natexlab{a}})}]{Shenker_BlackHolesButterfly_2014}%
  \BibitemOpen
  \bibfield  {author} {\bibinfo {author} {\bibfnamefont {S.~H.}\ \bibnamefont
  {Shenker}}\ and\ \bibinfo {author} {\bibfnamefont {D.}~\bibnamefont
  {Stanford}},\ }\bibfield  {title} {\enquote {\bibinfo {title} {Black holes
  and the butterfly effect},}\ }\href {\doibase 10.1007/jhep03(2014)067}
  {\bibfield  {journal} {\bibinfo  {journal} {Journal of High Energy Physics}\
  }\textbf {\bibinfo {volume} {2014}} (\bibinfo {year} {2014}{\natexlab{a}}),\
  10.1007/jhep03(2014)067}\BibitemShut {NoStop}%
\bibitem [{\citenamefont {Shenker}\ and\ \citenamefont
  {Stanford}(2014{\natexlab{b}})}]{Shenker_MultipleShocks_2014}%
  \BibitemOpen
  \bibfield  {author} {\bibinfo {author} {\bibfnamefont {S.~H.}\ \bibnamefont
  {Shenker}}\ and\ \bibinfo {author} {\bibfnamefont {D.}~\bibnamefont
  {Stanford}},\ }\bibfield  {title} {\enquote {\bibinfo {title} {Multiple
  shocks},}\ }\href {\doibase 10.1007/JHEP12(2014)046} {\bibfield  {journal}
  {\bibinfo  {journal} {Journal of High Energy Physics}\ }\textbf {\bibinfo
  {volume} {12}},\ \bibinfo {eid} {46} (\bibinfo {year}
  {2014}{\natexlab{b}})},\ \Eprint {http://arxiv.org/abs/1312.3296}
  {arXiv:1312.3296} \BibitemShut {NoStop}%
\bibitem [{\citenamefont
  {Hartnoll}(2015)}]{Hartnoll_TheoryUniversalIncoherent_2015}%
  \BibitemOpen
  \bibfield  {author} {\bibinfo {author} {\bibfnamefont {S.~A.}\ \bibnamefont
  {Hartnoll}},\ }\bibfield  {title} {\enquote {\bibinfo {title} {Theory of
  universal incoherent metallic transport},}\ }\href {\doibase
  10.1038/nphys3174} {\bibfield  {journal} {\bibinfo  {journal} {Nature
  Physics}\ }\textbf {\bibinfo {volume} {11}},\ \bibinfo {pages} {54} (\bibinfo
  {year} {2015})},\ \Eprint {http://arxiv.org/abs/1405.3651} {arXiv:1405.3651}
  \BibitemShut {NoStop}%
\bibitem [{\citenamefont {Shenker}\ and\ \citenamefont
  {Stanford}(2015)}]{Shenker_StringyEffectsScrambling_2015}%
  \BibitemOpen
  \bibfield  {author} {\bibinfo {author} {\bibfnamefont {S.~H.}\ \bibnamefont
  {Shenker}}\ and\ \bibinfo {author} {\bibfnamefont {D.}~\bibnamefont
  {Stanford}},\ }\bibfield  {title} {\enquote {\bibinfo {title} {Stringy
  effects in scrambling},}\ }\href {\doibase 10.1007/JHEP05(2015)132}
  {\bibfield  {journal} {\bibinfo  {journal} {Journal of High Energy Physics}\
  }\textbf {\bibinfo {volume} {5}},\ \bibinfo {eid} {132} (\bibinfo {year}
  {2015})},\ \Eprint {http://arxiv.org/abs/1412.6087} {arXiv:1412.6087}
  \BibitemShut {NoStop}%
\bibitem [{\citenamefont {Kitaev}(2015)}]{Kitaev_SimpleModelQuantum_2015}%
  \BibitemOpen
  \bibfield  {author} {\bibinfo {author} {\bibfnamefont {A.}~\bibnamefont
  {Kitaev}},\ }\href@noop {} {\enquote {\bibinfo {title} {A {{Simple Model}} of
  {{Quantum Holography}}},}\ } (\bibinfo {year} {2015}),\ \bibinfo {note}
  {{{KITP}} Strings Seminar and {{Entanglement}}}\BibitemShut {NoStop}%
\bibitem [{\citenamefont {Roberts}\ \emph {et~al.}(2015)\citenamefont
  {Roberts}, \citenamefont {Stanford},\ and\ \citenamefont
  {Susskind}}]{Roberts_LocalizedShocks_2015}%
  \BibitemOpen
  \bibfield  {author} {\bibinfo {author} {\bibfnamefont {D.~A.}\ \bibnamefont
  {Roberts}}, \bibinfo {author} {\bibfnamefont {D.}~\bibnamefont {Stanford}}, \
  and\ \bibinfo {author} {\bibfnamefont {L.}~\bibnamefont {Susskind}},\
  }\bibfield  {title} {\enquote {\bibinfo {title} {Localized shocks},}\ }\href
  {\doibase 10.1007/jhep03(2015)051} {\bibfield  {journal} {\bibinfo  {journal}
  {Journal of High Energy Physics}\ }\textbf {\bibinfo {volume} {2015}}
  (\bibinfo {year} {2015}),\ 10.1007/jhep03(2015)051}\BibitemShut {NoStop}%
\bibitem [{\citenamefont {Roberts}\ and\ \citenamefont
  {Stanford}(2015)}]{Roberts_DiagnosingChaosUsing_2015}%
  \BibitemOpen
  \bibfield  {author} {\bibinfo {author} {\bibfnamefont {D.~A.}\ \bibnamefont
  {Roberts}}\ and\ \bibinfo {author} {\bibfnamefont {D.}~\bibnamefont
  {Stanford}},\ }\bibfield  {title} {\enquote {\bibinfo {title} {Diagnosing
  {{Chaos Using Four}}-{{Point Functions}} in {{Two}}-{{Dimensional Conformal
  Field Theory}}},}\ }\href {\doibase 10.1103/PhysRevLett.115.131603}
  {\bibfield  {journal} {\bibinfo  {journal} {Physical Review Letters}\
  }\textbf {\bibinfo {volume} {115}},\ \bibinfo {eid} {131603} (\bibinfo {year}
  {2015})},\ \Eprint {http://arxiv.org/abs/1412.5123} {arXiv:1412.5123}
  \BibitemShut {NoStop}%
\bibitem [{\citenamefont {Maldacena}\ \emph {et~al.}(2016)\citenamefont
  {Maldacena}, \citenamefont {Shenker},\ and\ \citenamefont
  {Stanford}}]{Maldacena_BoundChaos_2016}%
  \BibitemOpen
  \bibfield  {author} {\bibinfo {author} {\bibfnamefont {J.}~\bibnamefont
  {Maldacena}}, \bibinfo {author} {\bibfnamefont {S.~H.}\ \bibnamefont
  {Shenker}}, \ and\ \bibinfo {author} {\bibfnamefont {D.}~\bibnamefont
  {Stanford}},\ }\bibfield  {title} {\enquote {\bibinfo {title} {A bound on
  chaos},}\ }\href {\doibase 10.1007/JHEP08(2016)106} {\bibfield  {journal}
  {\bibinfo  {journal} {Journal of High Energy Physics}\ }\textbf {\bibinfo
  {volume} {8}},\ \bibinfo {eid} {106} (\bibinfo {year} {2016})},\ \Eprint
  {http://arxiv.org/abs/1503.01409} {arXiv:1503.01409} \BibitemShut {NoStop}%
\bibitem [{\citenamefont {Aleiner}\ \emph {et~al.}(2016)\citenamefont
  {Aleiner}, \citenamefont {Faoro},\ and\ \citenamefont
  {Ioffe}}]{Aleiner_MicroscopicModelQuantum_2016}%
  \BibitemOpen
  \bibfield  {author} {\bibinfo {author} {\bibfnamefont {I.~L.}\ \bibnamefont
  {Aleiner}}, \bibinfo {author} {\bibfnamefont {L.}~\bibnamefont {Faoro}}, \
  and\ \bibinfo {author} {\bibfnamefont {L.~B.}\ \bibnamefont {Ioffe}},\
  }\bibfield  {title} {\enquote {\bibinfo {title} {Microscopic model of quantum
  butterfly effect: {{Out}}-of-time-order correlators and traveling combustion
  waves},}\ }\href {\doibase 10.1016/j.aop.2016.09.006} {\bibfield  {journal}
  {\bibinfo  {journal} {Annals of Physics}\ }\textbf {\bibinfo {volume}
  {375}},\ \bibinfo {pages} {378} (\bibinfo {year} {2016})},\ \Eprint
  {http://arxiv.org/abs/1609.01251} {arXiv:1609.01251} \BibitemShut {NoStop}%
\bibitem [{\citenamefont
  {Blake}(2016{\natexlab{a}})}]{Blake_UniversalChargeDiffusion_2016}%
  \BibitemOpen
  \bibfield  {author} {\bibinfo {author} {\bibfnamefont {M.}~\bibnamefont
  {Blake}},\ }\bibfield  {title} {\enquote {\bibinfo {title} {Universal
  {{Charge Diffusion}} and the {{Butterfly Effect}} in {{Holographic
  Theories}}},}\ }\href {\doibase 10.1103/PhysRevLett.117.091601} {\bibfield
  {journal} {\bibinfo  {journal} {Physical Review Letters}\ }\textbf {\bibinfo
  {volume} {117}},\ \bibinfo {pages} {091601} (\bibinfo {year}
  {2016}{\natexlab{a}})}\BibitemShut {NoStop}%
\bibitem [{\citenamefont
  {Blake}(2016{\natexlab{b}})}]{Blake_UniversalDiffusionIncoherent_2016}%
  \BibitemOpen
  \bibfield  {author} {\bibinfo {author} {\bibfnamefont {M.}~\bibnamefont
  {Blake}},\ }\bibfield  {title} {\enquote {\bibinfo {title} {Universal
  diffusion in incoherent black holes},}\ }\href {\doibase
  10.1103/PhysRevD.94.086014} {\bibfield  {journal} {\bibinfo  {journal}
  {Physical Review D}\ }\textbf {\bibinfo {volume} {94}},\ \bibinfo {pages}
  {086014} (\bibinfo {year} {2016}{\natexlab{b}})}\BibitemShut {NoStop}%
\bibitem [{\citenamefont
  {Chen}(2016)}]{Chen_UniversalLogarithmicScrambling_2016}%
  \BibitemOpen
  \bibfield  {author} {\bibinfo {author} {\bibfnamefont {Y.}~\bibnamefont
  {Chen}},\ }\bibfield  {title} {\enquote {\bibinfo {title} {Universal
  {{Logarithmic Scrambling}} in {{Many Body Localization}}},}\ }\href@noop {}
  {\bibfield  {journal} {\bibinfo  {journal} {ArXiv e-prints}\ } (\bibinfo
  {year} {2016})},\ \Eprint {http://arxiv.org/abs/1608.02765}
  {arXiv:1608.02765} \BibitemShut {NoStop}%
\bibitem [{\citenamefont {Lucas}\ and\ \citenamefont
  {Steinberg}(2016)}]{Lucas_ChargeDiffusionButterfly_2016}%
  \BibitemOpen
  \bibfield  {author} {\bibinfo {author} {\bibfnamefont {A.}~\bibnamefont
  {Lucas}}\ and\ \bibinfo {author} {\bibfnamefont {J.}~\bibnamefont
  {Steinberg}},\ }\bibfield  {title} {{\selectlanguage {en}\enquote {\bibinfo
  {title} {Charge diffusion and the butterfly effect in striped holographic
  matter},}\ }}\href {\doibase 10.1007/JHEP10(2016)143} {\bibfield  {journal}
  {\bibinfo  {journal} {Journal of High Energy Physics}\ }\textbf {\bibinfo
  {volume} {2016}},\ \bibinfo {pages} {143} (\bibinfo {year}
  {2016})}\BibitemShut {NoStop}%
\bibitem [{\citenamefont {Roberts}\ and\ \citenamefont
  {Swingle}(2016)}]{Roberts_LiebRobinsonBoundButterfly_2016}%
  \BibitemOpen
  \bibfield  {author} {\bibinfo {author} {\bibfnamefont {D.~A.}\ \bibnamefont
  {Roberts}}\ and\ \bibinfo {author} {\bibfnamefont {B.}~\bibnamefont
  {Swingle}},\ }\bibfield  {title} {\enquote {\bibinfo {title} {Lieb-{{Robinson
  Bound}} and the {{Butterfly Effect}} in {{Quantum Field Theories}}},}\ }\href
  {\doibase 10.1103/PhysRevLett.117.091602} {\bibfield  {journal} {\bibinfo
  {journal} {Physical Review Letters}\ }\textbf {\bibinfo {volume} {117}},\
  \bibinfo {pages} {091602} (\bibinfo {year} {2016})}\BibitemShut {NoStop}%
\bibitem [{\citenamefont {Hosur}\ \emph {et~al.}(2016)\citenamefont {Hosur},
  \citenamefont {Qi}, \citenamefont {Roberts},\ and\ \citenamefont
  {Yoshida}}]{Hosur_ChaosQuantumChannels_2016}%
  \BibitemOpen
  \bibfield  {author} {\bibinfo {author} {\bibfnamefont {P.}~\bibnamefont
  {Hosur}}, \bibinfo {author} {\bibfnamefont {X.-L.}\ \bibnamefont {Qi}},
  \bibinfo {author} {\bibfnamefont {D.~A.}\ \bibnamefont {Roberts}}, \ and\
  \bibinfo {author} {\bibfnamefont {B.}~\bibnamefont {Yoshida}},\ }\bibfield
  {title} {\enquote {\bibinfo {title} {Chaos in quantum channels},}\ }\href
  {\doibase 10.1007/JHEP02(2016)004} {\bibfield  {journal} {\bibinfo  {journal}
  {Journal of High Energy Physics}\ }\textbf {\bibinfo {volume} {2}},\ \bibinfo
  {eid} {4} (\bibinfo {year} {2016})},\ \Eprint
  {http://arxiv.org/abs/1511.04021} {arXiv:1511.04021} \BibitemShut {NoStop}%
\bibitem [{\citenamefont {Banerjee}\ and\ \citenamefont
  {Altman}(2017)}]{Banerjee_SolvableModelDynamical_2017}%
  \BibitemOpen
  \bibfield  {author} {\bibinfo {author} {\bibfnamefont {S.}~\bibnamefont
  {Banerjee}}\ and\ \bibinfo {author} {\bibfnamefont {E.}~\bibnamefont
  {Altman}},\ }\bibfield  {title} {\enquote {\bibinfo {title} {Solvable model
  for a dynamical quantum phase transition from fast to slow scrambling},}\
  }\href {\doibase 10.1103/PhysRevB.95.134302} {\bibfield  {journal} {\bibinfo
  {journal} {Physical Review B}\ }\textbf {\bibinfo {volume} {95}} (\bibinfo
  {year} {2017}),\ 10.1103/PhysRevB.95.134302}\BibitemShut {NoStop}%
\bibitem [{\citenamefont {Fan}\ \emph {et~al.}(2017)\citenamefont {Fan},
  \citenamefont {Zhang}, \citenamefont {Shen},\ and\ \citenamefont
  {Zhai}}]{Fan_OutoftimeorderCorrelationManybody_2017}%
  \BibitemOpen
  \bibfield  {author} {\bibinfo {author} {\bibfnamefont {R.}~\bibnamefont
  {Fan}}, \bibinfo {author} {\bibfnamefont {P.}~\bibnamefont {Zhang}}, \bibinfo
  {author} {\bibfnamefont {H.}~\bibnamefont {Shen}}, \ and\ \bibinfo {author}
  {\bibfnamefont {H.}~\bibnamefont {Zhai}},\ }\bibfield  {title} {\enquote
  {\bibinfo {title} {Out-of-time-order correlation for many-body
  localization},}\ }\href {\doibase 10.1016/j.scib.2017.04.011} {\bibfield
  {journal} {\bibinfo  {journal} {Science Bulletin}\ }\textbf {\bibinfo
  {volume} {62}},\ \bibinfo {pages} {707} (\bibinfo {year} {2017})}\BibitemShut
  {NoStop}%
\bibitem [{\citenamefont {Gu}\ \emph {et~al.}(2017)\citenamefont {Gu},
  \citenamefont {Qi},\ and\ \citenamefont
  {Stanford}}]{Gu_LocalCriticalityDiffusion_2017}%
  \BibitemOpen
  \bibfield  {author} {\bibinfo {author} {\bibfnamefont {Y.}~\bibnamefont
  {Gu}}, \bibinfo {author} {\bibfnamefont {X.-L.}\ \bibnamefont {Qi}}, \ and\
  \bibinfo {author} {\bibfnamefont {D.}~\bibnamefont {Stanford}},\ }\bibfield
  {title} {{\selectlanguage {en}\enquote {\bibinfo {title} {Local criticality,
  diffusion and chaos in generalized {{Sachdev}}-{{Ye}}-{{Kitaev}} models},}\
  }}\href {\doibase 10.1007/JHEP05(2017)125} {\bibfield  {journal} {\bibinfo
  {journal} {Journal of High Energy Physics}\ }\textbf {\bibinfo {volume}
  {2017}},\ \bibinfo {pages} {125} (\bibinfo {year} {2017})}\BibitemShut
  {NoStop}%
\bibitem [{\citenamefont {Roberts}\ and\ \citenamefont
  {Yoshida}(2017)}]{Roberts_ChaosComplexityDesign_2017}%
  \BibitemOpen
  \bibfield  {author} {\bibinfo {author} {\bibfnamefont {D.~A.}\ \bibnamefont
  {Roberts}}\ and\ \bibinfo {author} {\bibfnamefont {B.}~\bibnamefont
  {Yoshida}},\ }\bibfield  {title} {\enquote {\bibinfo {title} {Chaos and
  complexity by design},}\ }\href {\doibase 10.1007/JHEP04(2017)121} {\bibfield
   {journal} {\bibinfo  {journal} {Journal of High Energy Physics}\ }\textbf
  {\bibinfo {volume} {4}},\ \bibinfo {eid} {121} (\bibinfo {year} {2017})},\
  \Eprint {http://arxiv.org/abs/1610.04903} {arXiv:1610.04903} \BibitemShut
  {NoStop}%
\bibitem [{\citenamefont {Chen}\ and\ \citenamefont
  {Zhou}(2018)}]{Chen_OperatorScramblingQuantum_2018}%
  \BibitemOpen
  \bibfield  {author} {\bibinfo {author} {\bibfnamefont {X.}~\bibnamefont
  {Chen}}\ and\ \bibinfo {author} {\bibfnamefont {T.}~\bibnamefont {Zhou}},\
  }\bibfield  {title} {\enquote {\bibinfo {title} {Operator scrambling and
  quantum chaos},}\ }\href@noop {} {\bibfield  {journal} {\bibinfo  {journal}
  {ArXiv e-prints}\ } (\bibinfo {year} {2018})},\ \Eprint
  {http://arxiv.org/abs/1804.08655} {arXiv:1804.08655} \BibitemShut {NoStop}%
\bibitem [{\citenamefont {Huang}\ \emph {et~al.}(2017)\citenamefont {Huang},
  \citenamefont {Zhang},\ and\ \citenamefont
  {Chen}}]{Huang_OutoftimeorderedCorrelatorsManybody_2017}%
  \BibitemOpen
  \bibfield  {author} {\bibinfo {author} {\bibfnamefont {Y.}~\bibnamefont
  {Huang}}, \bibinfo {author} {\bibfnamefont {Y.-L.}\ \bibnamefont {Zhang}}, \
  and\ \bibinfo {author} {\bibfnamefont {X.}~\bibnamefont {Chen}},\ }\bibfield
  {title} {\enquote {\bibinfo {title} {Out-of-time-ordered correlators in
  many-body localized systems},}\ }\href {\doibase 10.1002/andp.201600318}
  {\bibfield  {journal} {\bibinfo  {journal} {Annalen der Physik}\ }\textbf
  {\bibinfo {volume} {529}},\ \bibinfo {pages} {1600318} (\bibinfo {year}
  {2017})},\ \Eprint {http://arxiv.org/abs/1608.01091} {arXiv:1608.01091}
  \BibitemShut {NoStop}%
\bibitem [{\citenamefont {Iyoda}\ and\ \citenamefont
  {Sagawa}(2018)}]{Iyoda_ScramblingQuantumInformation_2018}%
  \BibitemOpen
  \bibfield  {author} {\bibinfo {author} {\bibfnamefont {E.}~\bibnamefont
  {Iyoda}}\ and\ \bibinfo {author} {\bibfnamefont {T.}~\bibnamefont {Sagawa}},\
  }\bibfield  {title} {\enquote {\bibinfo {title} {Scrambling of {{Quantum
  Information}} in {{Quantum Many}}-{{Body Systems}}},}\ }\href {\doibase
  10.1103/PhysRevA.97.042330} {\bibfield  {journal} {\bibinfo  {journal}
  {Physical Review A}\ }\textbf {\bibinfo {volume} {97}},\ \bibinfo {pages}
  {042330} (\bibinfo {year} {2018})},\ \Eprint
  {http://arxiv.org/abs/1704.04850} {arXiv:1704.04850} \BibitemShut {NoStop}%
\bibitem [{\citenamefont {Yoshida}\ and\ \citenamefont
  {Kitaev}(2017)}]{Yoshida_EfficientDecodingHaydenPreskill_2017}%
  \BibitemOpen
  \bibfield  {author} {\bibinfo {author} {\bibfnamefont {B.}~\bibnamefont
  {Yoshida}}\ and\ \bibinfo {author} {\bibfnamefont {A.}~\bibnamefont
  {Kitaev}},\ }\bibfield  {title} {\enquote {\bibinfo {title} {Efficient
  decoding for the {{Hayden}}-{{Preskill}} protocol},}\ }\href@noop {}
  {\bibfield  {journal} {\bibinfo  {journal} {ArXiv e-prints}\ } (\bibinfo
  {year} {2017})},\ \Eprint {http://arxiv.org/abs/1710.03363}
  {arXiv:1710.03363} \BibitemShut {NoStop}%
\bibitem [{\citenamefont {Lin}\ and\ \citenamefont
  {Motrunich}(2018)}]{Lin_OutoftimeorderedCorrelatorsQuantum_2018}%
  \BibitemOpen
  \bibfield  {author} {\bibinfo {author} {\bibfnamefont {C.-J.}\ \bibnamefont
  {Lin}}\ and\ \bibinfo {author} {\bibfnamefont {O.~I.}\ \bibnamefont
  {Motrunich}},\ }\bibfield  {title} {\enquote {\bibinfo {title}
  {Out-of-time-ordered correlators in a quantum {{Ising}} chain},}\ }\href
  {\doibase 10.1103/PhysRevB.97.144304} {\bibfield  {journal} {\bibinfo
  {journal} {Physical Review B}\ }\textbf {\bibinfo {volume} {97}},\ \bibinfo
  {pages} {144304} (\bibinfo {year} {2018})}\BibitemShut {NoStop}%
\bibitem [{\citenamefont {Pappalardi}\ \emph {et~al.}(2018)\citenamefont
  {Pappalardi}, \citenamefont {Russomanno}, \citenamefont {{\v Z}unkovi{\v c}},
  \citenamefont {Iemini}, \citenamefont {Silva},\ and\ \citenamefont
  {Fazio}}]{Pappalardi_ScramblingEntanglementSpreading_2018}%
  \BibitemOpen
  \bibfield  {author} {\bibinfo {author} {\bibfnamefont {S.}~\bibnamefont
  {Pappalardi}}, \bibinfo {author} {\bibfnamefont {A.}~\bibnamefont
  {Russomanno}}, \bibinfo {author} {\bibfnamefont {B.}~\bibnamefont {{\v
  Z}unkovi{\v c}}}, \bibinfo {author} {\bibfnamefont {F.}~\bibnamefont
  {Iemini}}, \bibinfo {author} {\bibfnamefont {A.}~\bibnamefont {Silva}}, \
  and\ \bibinfo {author} {\bibfnamefont {R.}~\bibnamefont {Fazio}},\ }\bibfield
   {title} {\enquote {\bibinfo {title} {Scrambling and entanglement spreading
  in long-range spin chains},}\ }\href {\doibase 10.1103/PhysRevB.98.134303}
  {\bibfield  {journal} {\bibinfo  {journal} {Physical Review B}\ }\textbf
  {\bibinfo {volume} {98}},\ \bibinfo {pages} {134303} (\bibinfo {year}
  {2018})},\ \Eprint {http://arxiv.org/abs/1806.00022} {arXiv:1806.00022}
  \BibitemShut {NoStop}%
\bibitem [{\citenamefont {Yunger~Halpern}\ \emph {et~al.}(2019)\citenamefont
  {Yunger~Halpern}, \citenamefont {Bartolotta},\ and\ \citenamefont
  {Pollack}}]{YungerHalpern_EntropicUncertaintyRelations_2019}%
  \BibitemOpen
  \bibfield  {author} {\bibinfo {author} {\bibfnamefont {N.}~\bibnamefont
  {Yunger~Halpern}}, \bibinfo {author} {\bibfnamefont {A.}~\bibnamefont
  {Bartolotta}}, \ and\ \bibinfo {author} {\bibfnamefont {J.}~\bibnamefont
  {Pollack}},\ }\bibfield  {title} {{\selectlanguage {en}\enquote {\bibinfo
  {title} {Entropic uncertainty relations for quantum information
  scrambling},}\ }}\href {\doibase 10.1038/s42005-019-0179-8} {\bibfield
  {journal} {\bibinfo  {journal} {Communications Physics}\ }\textbf {\bibinfo
  {volume} {2}},\ \bibinfo {pages} {1} (\bibinfo {year} {2019})}\BibitemShut
  {NoStop}%
\bibitem [{\citenamefont {Vermersch}\ \emph {et~al.}(2019)\citenamefont
  {Vermersch}, \citenamefont {Elben}, \citenamefont {Sieberer}, \citenamefont
  {Yao},\ and\ \citenamefont {Zoller}}]{Vermersch_2019}%
  \BibitemOpen
  \bibfield  {author} {\bibinfo {author} {\bibfnamefont {B.}~\bibnamefont
  {Vermersch}}, \bibinfo {author} {\bibfnamefont {A.}~\bibnamefont {Elben}},
  \bibinfo {author} {\bibfnamefont {L.~M.}\ \bibnamefont {Sieberer}}, \bibinfo
  {author} {\bibfnamefont {N.~Y.}\ \bibnamefont {Yao}}, \ and\ \bibinfo
  {author} {\bibfnamefont {P.}~\bibnamefont {Zoller}},\ }\bibfield  {title}
  {\enquote {\bibinfo {title} {Probing scrambling using statistical
  correlations between randomized measurements},}\ }\href {\doibase
  10.1103/PhysRevX.9.021061} {\bibfield  {journal} {\bibinfo  {journal} {Phys.
  Rev. X}\ }\textbf {\bibinfo {volume} {9}},\ \bibinfo {pages} {021061}
  (\bibinfo {year} {2019})}\BibitemShut {NoStop}%
\bibitem [{\citenamefont {Harrow}\ \emph {et~al.}(2021)\citenamefont {Harrow},
  \citenamefont {Kong}, \citenamefont {Liu}, \citenamefont {Mehraban},\ and\
  \citenamefont {Shor}}]{Harrow21}%
  \BibitemOpen
  \bibfield  {author} {\bibinfo {author} {\bibfnamefont {A.~W.}\ \bibnamefont
  {Harrow}}, \bibinfo {author} {\bibfnamefont {L.}~\bibnamefont {Kong}},
  \bibinfo {author} {\bibfnamefont {Z.-W.}\ \bibnamefont {Liu}}, \bibinfo
  {author} {\bibfnamefont {S.}~\bibnamefont {Mehraban}}, \ and\ \bibinfo
  {author} {\bibfnamefont {P.~W.}\ \bibnamefont {Shor}},\ }\bibfield  {title}
  {\enquote {\bibinfo {title} {Separation of out-of-time-ordered correlation
  and entanglement},}\ }\href {\doibase 10.1103/PRXQuantum.2.020339} {\bibfield
   {journal} {\bibinfo  {journal} {PRX Quantum}\ }\textbf {\bibinfo {volume}
  {2}},\ \bibinfo {pages} {020339} (\bibinfo {year} {2021})}\BibitemShut
  {NoStop}%
\bibitem [{\citenamefont {Cao}\ \emph {et~al.}(2021)\citenamefont {Cao},
  \citenamefont {Xu},\ and\ \citenamefont {del Campo}}]{Cao2021}%
  \BibitemOpen
  \bibfield  {author} {\bibinfo {author} {\bibfnamefont {Z.}~\bibnamefont
  {Cao}}, \bibinfo {author} {\bibfnamefont {Z.}~\bibnamefont {Xu}}, \ and\
  \bibinfo {author} {\bibfnamefont {A.}~\bibnamefont {del Campo}},\ }\bibfield
  {title} {\enquote {\bibinfo {title} {Diagnosing quantum chaos in multipartite
  systems},}\ }\href@noop {} {\bibfield  {journal} {\bibinfo  {journal} {arXiv
  preprint arXiv:2111.12475}\ } (\bibinfo {year} {2021})},\ \Eprint
  {http://arxiv.org/abs/2111.12475} {2111.12475} \BibitemShut {NoStop}%
\bibitem [{\citenamefont {Braum\"{u}ller}\ \emph {et~al.}(2021)\citenamefont
  {Braum\"{u}ller}, \citenamefont {Karamlou}, \citenamefont {Yanay},
  \citenamefont {Kannan}, \citenamefont {Kim}, \citenamefont {Kjaergaard},
  \citenamefont {Melville}, \citenamefont {Niedzielski}, \citenamefont {Sung},
  \citenamefont {Veps\"{a}l\"{a}inen}, \citenamefont {Winik}, \citenamefont
  {Yoder}, \citenamefont {Orlando}, \citenamefont {Gustavsson}, \citenamefont
  {Tahan},\ and\ \citenamefont {Oliver}}]{Braumller2021}%
  \BibitemOpen
  \bibfield  {author} {\bibinfo {author} {\bibfnamefont {J.}~\bibnamefont
  {Braum\"{u}ller}}, \bibinfo {author} {\bibfnamefont {A.~H.}\ \bibnamefont
  {Karamlou}}, \bibinfo {author} {\bibfnamefont {Y.}~\bibnamefont {Yanay}},
  \bibinfo {author} {\bibfnamefont {B.}~\bibnamefont {Kannan}}, \bibinfo
  {author} {\bibfnamefont {D.}~\bibnamefont {Kim}}, \bibinfo {author}
  {\bibfnamefont {M.}~\bibnamefont {Kjaergaard}}, \bibinfo {author}
  {\bibfnamefont {A.}~\bibnamefont {Melville}}, \bibinfo {author}
  {\bibfnamefont {B.~M.}\ \bibnamefont {Niedzielski}}, \bibinfo {author}
  {\bibfnamefont {Y.}~\bibnamefont {Sung}}, \bibinfo {author} {\bibfnamefont
  {A.}~\bibnamefont {Veps\"{a}l\"{a}inen}}, \bibinfo {author} {\bibfnamefont
  {R.}~\bibnamefont {Winik}}, \bibinfo {author} {\bibfnamefont {J.~L.}\
  \bibnamefont {Yoder}}, \bibinfo {author} {\bibfnamefont {T.~P.}\ \bibnamefont
  {Orlando}}, \bibinfo {author} {\bibfnamefont {S.}~\bibnamefont {Gustavsson}},
  \bibinfo {author} {\bibfnamefont {C.}~\bibnamefont {Tahan}}, \ and\ \bibinfo
  {author} {\bibfnamefont {W.~D.}\ \bibnamefont {Oliver}},\ }\bibfield  {title}
  {\enquote {\bibinfo {title} {Probing quantum information propagation with
  out-of-time-ordered correlators},}\ }\href {\doibase
  10.1038/s41567-021-01430-w} {\bibfield  {journal} {\bibinfo  {journal}
  {Nature Physics}\ } (\bibinfo {year} {2021}),\
  10.1038/s41567-021-01430-w}\BibitemShut {NoStop}%
\bibitem [{\citenamefont {Mi}\ \emph {et~al.}(2021)\citenamefont {Mi},
  \citenamefont {Roushan}, \citenamefont {Quintana}, \citenamefont {Mandrà},
  \citenamefont {Marshall}, \citenamefont {Neill}, \citenamefont {Arute},
  \citenamefont {Arya}, \citenamefont {Atalaya}, \citenamefont {Babbush},\ and\
  \citenamefont {et~al.}}]{Mi2021}%
  \BibitemOpen
  \bibfield  {author} {\bibinfo {author} {\bibfnamefont {X.}~\bibnamefont
  {Mi}}, \bibinfo {author} {\bibfnamefont {P.}~\bibnamefont {Roushan}},
  \bibinfo {author} {\bibfnamefont {C.}~\bibnamefont {Quintana}}, \bibinfo
  {author} {\bibfnamefont {S.}~\bibnamefont {Mandrà}}, \bibinfo {author}
  {\bibfnamefont {J.}~\bibnamefont {Marshall}}, \bibinfo {author}
  {\bibfnamefont {C.}~\bibnamefont {Neill}}, \bibinfo {author} {\bibfnamefont
  {F.}~\bibnamefont {Arute}}, \bibinfo {author} {\bibfnamefont
  {K.}~\bibnamefont {Arya}}, \bibinfo {author} {\bibfnamefont {J.}~\bibnamefont
  {Atalaya}}, \bibinfo {author} {\bibfnamefont {R.}~\bibnamefont {Babbush}}, \
  and\ \bibinfo {author} {\bibnamefont {et~al.}},\ }\bibfield  {title}
  {\enquote {\bibinfo {title} {Information scrambling in quantum circuits},}\
  }\href {\doibase 10.1126/science.abg5029} {\bibfield  {journal} {\bibinfo
  {journal} {Science}\ }\textbf {\bibinfo {volume} {374}},\ \bibinfo {pages}
  {1479–1483} (\bibinfo {year} {2021})}\BibitemShut {NoStop}%
\bibitem [{\citenamefont {Lerose}\ and\ \citenamefont
  {Pappalardi}(2020{\natexlab{a}})}]{Lerose20}%
  \BibitemOpen
  \bibfield  {author} {\bibinfo {author} {\bibfnamefont {A.}~\bibnamefont
  {Lerose}}\ and\ \bibinfo {author} {\bibfnamefont {S.}~\bibnamefont
  {Pappalardi}},\ }\bibfield  {title} {\enquote {\bibinfo {title} {Bridging
  entanglement dynamics and chaos in semiclassical systems},}\ }\href {\doibase
  10.1103/PhysRevA.102.032404} {\bibfield  {journal} {\bibinfo  {journal}
  {Phys. Rev. A}\ }\textbf {\bibinfo {volume} {102}},\ \bibinfo {pages}
  {032404} (\bibinfo {year} {2020}{\natexlab{a}})}\BibitemShut {NoStop}%
\bibitem [{\citenamefont {Gonz{\'a}lez~Alonso}\ \emph
  {et~al.}(2019)\citenamefont {Gonz{\'a}lez~Alonso}, \citenamefont
  {Yunger~Halpern},\ and\ \citenamefont
  {Dressel}}]{GonzalezAlonso_OutofTimeOrderedCorrelatorQuasiprobabilitiesRobustly_2019}%
  \BibitemOpen
  \bibfield  {author} {\bibinfo {author} {\bibfnamefont {J.~R.}\ \bibnamefont
  {Gonz{\'a}lez~Alonso}}, \bibinfo {author} {\bibfnamefont {N.}~\bibnamefont
  {Yunger~Halpern}}, \ and\ \bibinfo {author} {\bibfnamefont {J.}~\bibnamefont
  {Dressel}},\ }\bibfield  {title} {{\selectlanguage {en}\enquote {\bibinfo
  {title} {Out-of-{{Time}}-{{Ordered}}-{{Correlator Quasiprobabilities Robustly
  Witness Scrambling}}},}\ }}\href {\doibase 10.1103/PhysRevLett.122.040404}
  {\bibfield  {journal} {\bibinfo  {journal} {Physical Review Letters}\
  }\textbf {\bibinfo {volume} {122}},\ \bibinfo {pages} {040404} (\bibinfo
  {year} {2019})}\BibitemShut {NoStop}%
\bibitem [{\citenamefont {Cerf}\ and\ \citenamefont
  {Adami}(1998)}]{Cerf_InformationTheoryQuantum_1998}%
  \BibitemOpen
  \bibfield  {author} {\bibinfo {author} {\bibfnamefont {N.~J.}\ \bibnamefont
  {Cerf}}\ and\ \bibinfo {author} {\bibfnamefont {C.}~\bibnamefont {Adami}},\
  }\bibfield  {title} {{\selectlanguage {en}\enquote {\bibinfo {title}
  {Information theory of quantum entanglement and measurement},}\ }}\href
  {\doibase 10.1016/S0167-2789(98)00045-1} {\bibfield  {journal} {\bibinfo
  {journal} {Physica D: Nonlinear Phenomena}\ }\textbf {\bibinfo {volume}
  {120}},\ \bibinfo {pages} {62} (\bibinfo {year} {1998})}\BibitemShut
  {NoStop}%
\bibitem [{\citenamefont {Wang}\ \emph {et~al.}(2004)\citenamefont {Wang},
  \citenamefont {Ghose}, \citenamefont {Sanders},\ and\ \citenamefont
  {Hu}}]{Wang_EntanglementSignatureQuantum_2004}%
  \BibitemOpen
  \bibfield  {author} {\bibinfo {author} {\bibfnamefont {X.}~\bibnamefont
  {Wang}}, \bibinfo {author} {\bibfnamefont {S.}~\bibnamefont {Ghose}},
  \bibinfo {author} {\bibfnamefont {B.~C.}\ \bibnamefont {Sanders}}, \ and\
  \bibinfo {author} {\bibfnamefont {B.}~\bibnamefont {Hu}},\ }\bibfield
  {title} {\enquote {\bibinfo {title} {Entanglement as a signature of quantum
  chaos},}\ }\href {\doibase 10.1103/physreve.70.016217} {\bibfield  {journal}
  {\bibinfo  {journal} {Physical Review E}\ }\textbf {\bibinfo {volume} {70}},\
  \bibinfo {pages} {016217} (\bibinfo {year} {2004})}\BibitemShut {NoStop}%
\bibitem [{\citenamefont {Atkins}\ and\ \citenamefont
  {Dobson}(1971)}]{P.W.Atkins_AngularMomentumCoherent_1971}%
  \BibitemOpen
  \bibfield  {author} {\bibinfo {author} {\bibfnamefont {P.~W.}\ \bibnamefont
  {Atkins}}\ and\ \bibinfo {author} {\bibfnamefont {J.~C.}\ \bibnamefont
  {Dobson}},\ }\bibfield  {title} {{\selectlanguage {en}\enquote {\bibinfo
  {title} {Angular {{Momentum Coherent States}}},}\ }}\href@noop {} {\bibfield
  {journal} {\bibinfo  {journal} {Proceedings of the Royal Society of London.
  Series A, Mathematical and Physical Sciences}\ }\textbf {\bibinfo {volume}
  {321}},\ \bibinfo {pages} {321} (\bibinfo {year} {1971})}\BibitemShut
  {NoStop}%
\bibitem [{\citenamefont {Arecchi}\ \emph {et~al.}(1972)\citenamefont
  {Arecchi}, \citenamefont {Courtens}, \citenamefont {Gilmore},\ and\
  \citenamefont {Thomas}}]{Arecchi_AtomicCoherentStates_1972}%
  \BibitemOpen
  \bibfield  {author} {\bibinfo {author} {\bibfnamefont {F.~T.}\ \bibnamefont
  {Arecchi}}, \bibinfo {author} {\bibfnamefont {E.}~\bibnamefont {Courtens}},
  \bibinfo {author} {\bibfnamefont {R.}~\bibnamefont {Gilmore}}, \ and\
  \bibinfo {author} {\bibfnamefont {H.}~\bibnamefont {Thomas}},\ }\bibfield
  {title} {\enquote {\bibinfo {title} {Atomic {{Coherent States}} in {{Quantum
  Optics}}},}\ }\href {\doibase 10.1103/physreva.6.2211} {\bibfield  {journal}
  {\bibinfo  {journal} {Physical Review A}\ }\textbf {\bibinfo {volume} {6}},\
  \bibinfo {pages} {2211} (\bibinfo {year} {1972})}\BibitemShut {NoStop}%
\bibitem [{\citenamefont {Zhang}\ \emph {et~al.}(1990)\citenamefont {Zhang},
  \citenamefont {Feng},\ and\ \citenamefont
  {Gilmore}}]{Zhang_CoherentStatesTheory_1990}%
  \BibitemOpen
  \bibfield  {author} {\bibinfo {author} {\bibfnamefont {W.-M.}\ \bibnamefont
  {Zhang}}, \bibinfo {author} {\bibfnamefont {D.~H.}\ \bibnamefont {Feng}}, \
  and\ \bibinfo {author} {\bibfnamefont {R.}~\bibnamefont {Gilmore}},\
  }\bibfield  {title} {\enquote {\bibinfo {title} {Coherent states: {{Theory}}
  and some applications},}\ }\href {\doibase 10.1103/RevModPhys.62.867}
  {\bibfield  {journal} {\bibinfo  {journal} {Reviews of Modern Physics}\
  }\textbf {\bibinfo {volume} {62}},\ \bibinfo {pages} {867} (\bibinfo {year}
  {1990})}\BibitemShut {NoStop}%
\bibitem [{\citenamefont {Haake}\ \emph {et~al.}(1987)\citenamefont {Haake},
  \citenamefont {Ku{\'s}},\ and\ \citenamefont
  {Scharf}}]{Haake_ClassicalQuantumChaos_1987}%
  \BibitemOpen
  \bibfield  {author} {\bibinfo {author} {\bibfnamefont {F.}~\bibnamefont
  {Haake}}, \bibinfo {author} {\bibfnamefont {M.}~\bibnamefont {Ku{\'s}}}, \
  and\ \bibinfo {author} {\bibfnamefont {R.}~\bibnamefont {Scharf}},\
  }\bibfield  {title} {\enquote {\bibinfo {title} {Classical and quantum chaos
  for a kicked top},}\ }\href {\doibase 10.1007/BF01303727} {\bibfield
  {journal} {\bibinfo  {journal} {Zeitschrift f\"ur Physik B Condensed Matter}\
  }\textbf {\bibinfo {volume} {65}},\ \bibinfo {pages} {381} (\bibinfo {year}
  {1987})}\BibitemShut {NoStop}%
\bibitem [{\citenamefont {Ghose}\ \emph
  {et~al.}(2008{\natexlab{a}})\citenamefont {Ghose}, \citenamefont {Paul},\
  and\ \citenamefont {Stock}}]{Ghose_QuantumChaosTunneling_2008}%
  \BibitemOpen
  \bibfield  {author} {\bibinfo {author} {\bibfnamefont {S.}~\bibnamefont
  {Ghose}}, \bibinfo {author} {\bibfnamefont {C.~R.}\ \bibnamefont {Paul}}, \
  and\ \bibinfo {author} {\bibfnamefont {R.}~\bibnamefont {Stock}},\ }\bibfield
   {title} {\enquote {\bibinfo {title} {Quantum chaos and tunneling in the
  kicked top},}\ }\href {\doibase 10.1134/s1054660x0809017x} {\bibfield
  {journal} {\bibinfo  {journal} {Laser Physics}\ }\textbf {\bibinfo {volume}
  {18}},\ \bibinfo {pages} {1098} (\bibinfo {year}
  {2008}{\natexlab{a}})}\BibitemShut {NoStop}%
\bibitem [{\citenamefont {Ghose}\ \emph
  {et~al.}(2008{\natexlab{b}})\citenamefont {Ghose}, \citenamefont {Stock},
  \citenamefont {Jessen}, \citenamefont {Lal},\ and\ \citenamefont
  {Silberfarb}}]{Ghose_ChaosEntanglementDecoherence_2008}%
  \BibitemOpen
  \bibfield  {author} {\bibinfo {author} {\bibfnamefont {S.}~\bibnamefont
  {Ghose}}, \bibinfo {author} {\bibfnamefont {R.}~\bibnamefont {Stock}},
  \bibinfo {author} {\bibfnamefont {P.}~\bibnamefont {Jessen}}, \bibinfo
  {author} {\bibfnamefont {R.}~\bibnamefont {Lal}}, \ and\ \bibinfo {author}
  {\bibfnamefont {A.}~\bibnamefont {Silberfarb}},\ }\bibfield  {title}
  {{\selectlanguage {en}\enquote {\bibinfo {title} {Chaos, entanglement, and
  decoherence in the quantum kicked top},}\ }}\href {\doibase
  10.1103/PhysRevA.78.042318} {\bibfield  {journal} {\bibinfo  {journal}
  {Physical Review A}\ }\textbf {\bibinfo {volume} {78}} (\bibinfo {year}
  {2008}{\natexlab{b}}),\ 10.1103/PhysRevA.78.042318}\BibitemShut {NoStop}%
\bibitem [{\citenamefont {Kumari}\ and\ \citenamefont
  {Ghose}(2019)}]{Kumari_UntanglingEntanglementChaos_2019}%
  \BibitemOpen
  \bibfield  {author} {\bibinfo {author} {\bibfnamefont {M.}~\bibnamefont
  {Kumari}}\ and\ \bibinfo {author} {\bibfnamefont {S.}~\bibnamefont {Ghose}},\
  }\bibfield  {title} {\enquote {\bibinfo {title} {Untangling entanglement and
  chaos},}\ }\href {\doibase 10.1103/PhysRevA.99.042311} {\bibfield  {journal}
  {\bibinfo  {journal} {Physical Review A}\ }\textbf {\bibinfo {volume} {99}},\
  \bibinfo {pages} {042311} (\bibinfo {year} {2019})}\BibitemShut {NoStop}%
\bibitem [{\citenamefont {Haake}(2001)}]{Haake_QuantumSignaturesChaos_2001}%
  \BibitemOpen
  \bibfield  {author} {\bibinfo {author} {\bibfnamefont {F.}~\bibnamefont
  {Haake}},\ }\href {\doibase 10.1007/978-3-662-04506-0} {\emph {\bibinfo
  {title} {Quantum {{Signatures}} of {{Chaos}}}}}\ (\bibinfo  {publisher}
  {{Springer Berlin Heidelberg}},\ \bibinfo {year} {2001})\BibitemShut
  {NoStop}%
\bibitem [{\citenamefont {Husimi}(1940)}]{Husimi_FormalPropertiesDensity_1940}%
  \BibitemOpen
  \bibfield  {author} {\bibinfo {author} {\bibfnamefont {K.}~\bibnamefont
  {Husimi}},\ }\bibfield  {title} {\enquote {\bibinfo {title} {Some {{Formal
  Properties}} of the {{Density Matrix}}},}\ }\href {\doibase
  10.11429/ppmsj1919.22.4_264} {\bibfield  {journal} {\bibinfo  {journal}
  {Proceedings of the Physico-Mathematical Society of Japan. 3rd Series}\
  }\textbf {\bibinfo {volume} {22}},\ \bibinfo {pages} {264} (\bibinfo {year}
  {1940})}\BibitemShut {NoStop}%
\bibitem [{\citenamefont {Seshadri}\ \emph {et~al.}(2018)\citenamefont
  {Seshadri}, \citenamefont {Madhok},\ and\ \citenamefont
  {Lakshminarayan}}]{Seshadri_TripartiteMutualInformation_2018}%
  \BibitemOpen
  \bibfield  {author} {\bibinfo {author} {\bibfnamefont {A.}~\bibnamefont
  {Seshadri}}, \bibinfo {author} {\bibfnamefont {V.}~\bibnamefont {Madhok}}, \
  and\ \bibinfo {author} {\bibfnamefont {A.}~\bibnamefont {Lakshminarayan}},\
  }\bibfield  {title} {\enquote {\bibinfo {title} {Tripartite mutual
  information, entanglement, and scrambling in permutation symmetric systems
  with an application to quantum chaos},}\ }\href {\doibase
  10.1103/PhysRevE.98.052205} {\bibfield  {journal} {\bibinfo  {journal}
  {Physical Review E}\ }\textbf {\bibinfo {volume} {98}},\ \bibinfo {pages}
  {052205} (\bibinfo {year} {2018})},\ \Eprint
  {http://arxiv.org/abs/1806.00113} {arXiv:1806.00113} \BibitemShut {NoStop}%
\bibitem [{\citenamefont {Shammah}\ \emph {et~al.}(2018)\citenamefont
  {Shammah}, \citenamefont {Ahmed}, \citenamefont {Lambert}, \citenamefont
  {De~Liberato},\ and\ \citenamefont {Nori}}]{Shammah_OpenQuantumSystems_2018}%
  \BibitemOpen
  \bibfield  {author} {\bibinfo {author} {\bibfnamefont {N.}~\bibnamefont
  {Shammah}}, \bibinfo {author} {\bibfnamefont {S.}~\bibnamefont {Ahmed}},
  \bibinfo {author} {\bibfnamefont {N.}~\bibnamefont {Lambert}}, \bibinfo
  {author} {\bibfnamefont {S.}~\bibnamefont {De~Liberato}}, \ and\ \bibinfo
  {author} {\bibfnamefont {F.}~\bibnamefont {Nori}},\ }\bibfield  {title}
  {\enquote {\bibinfo {title} {Open quantum systems with local and collective
  incoherent processes: {{Efficient}} numerical simulation using permutational
  invariance},}\ }\href {\doibase 10.1103/PhysRevA.98.063815} {\bibfield
  {journal} {\bibinfo  {journal} {Physical Review A}\ }\textbf {\bibinfo
  {volume} {98}} (\bibinfo {year} {2018}),\ 10.1103/PhysRevA.98.063815},\
  \Eprint {http://arxiv.org/abs/1805.05129} {arXiv:1805.05129} \BibitemShut
  {NoStop}%
\bibitem [{\citenamefont {Johansson}\ \emph {et~al.}(2012)\citenamefont
  {Johansson}, \citenamefont {Nation},\ and\ \citenamefont
  {Nori}}]{Johansson_QuTiPOpensourcePython_2012}%
  \BibitemOpen
  \bibfield  {author} {\bibinfo {author} {\bibfnamefont {J.~R.}\ \bibnamefont
  {Johansson}}, \bibinfo {author} {\bibfnamefont {P.~D.}\ \bibnamefont
  {Nation}}, \ and\ \bibinfo {author} {\bibfnamefont {F.}~\bibnamefont
  {Nori}},\ }\bibfield  {title} {\enquote {\bibinfo {title} {{{QuTiP}}: {{An}}
  open-source {{Python}} framework for the dynamics of open quantum systems},}\
  }\href {\doibase 10.1016/j.cpc.2012.02.021} {\bibfield  {journal} {\bibinfo
  {journal} {Computer Physics Communications}\ }\textbf {\bibinfo {volume}
  {183}},\ \bibinfo {pages} {1760} (\bibinfo {year} {2012})},\ \Eprint
  {http://arxiv.org/abs/1110.0573} {arXiv:1110.0573} \BibitemShut {NoStop}%
\bibitem [{\citenamefont {Johansson}\ \emph {et~al.}(2013)\citenamefont
  {Johansson}, \citenamefont {Nation},\ and\ \citenamefont
  {Nori}}]{Johansson_QuTiPPythonFramework_2013}%
  \BibitemOpen
  \bibfield  {author} {\bibinfo {author} {\bibfnamefont {J.~R.}\ \bibnamefont
  {Johansson}}, \bibinfo {author} {\bibfnamefont {P.~D.}\ \bibnamefont
  {Nation}}, \ and\ \bibinfo {author} {\bibfnamefont {F.}~\bibnamefont
  {Nori}},\ }\bibfield  {title} {\enquote {\bibinfo {title} {{{QuTiP}} 2: {{A
  Python}} framework for the dynamics of open quantum systems},}\ }\href
  {\doibase 10.1016/j.cpc.2012.11.019} {\bibfield  {journal} {\bibinfo
  {journal} {Computer Physics Communications}\ }\textbf {\bibinfo {volume}
  {184}},\ \bibinfo {pages} {1234} (\bibinfo {year} {2013})},\ \Eprint
  {http://arxiv.org/abs/1211.6518} {arXiv:1211.6518} \BibitemShut {NoStop}%
\bibitem [{\citenamefont {Neill}\ \emph {et~al.}(2016)\citenamefont {Neill},
  \citenamefont {Roushan}, \citenamefont {Fang}, \citenamefont {Chen},
  \citenamefont {Kolodrubetz}, \citenamefont {Chen}, \citenamefont {Megrant},
  \citenamefont {Barends}, \citenamefont {Campbell}, \citenamefont {Chiaro},
  \citenamefont {Dunsworth}, \citenamefont {Jeffrey}, \citenamefont {Kelly},
  \citenamefont {Mutus}, \citenamefont {O'Malley}, \citenamefont {Quintana},
  \citenamefont {Sank}, \citenamefont {Vainsencher}, \citenamefont {Wenner},
  \citenamefont {White}, \citenamefont {Polkovnikov},\ and\ \citenamefont
  {Martinis}}]{Neill_ErgodicDynamicsThermalization_2016}%
  \BibitemOpen
  \bibfield  {author} {\bibinfo {author} {\bibfnamefont {C.}~\bibnamefont
  {Neill}}, \bibinfo {author} {\bibfnamefont {P.}~\bibnamefont {Roushan}},
  \bibinfo {author} {\bibfnamefont {M.}~\bibnamefont {Fang}}, \bibinfo {author}
  {\bibfnamefont {Y.}~\bibnamefont {Chen}}, \bibinfo {author} {\bibfnamefont
  {M.}~\bibnamefont {Kolodrubetz}}, \bibinfo {author} {\bibfnamefont
  {Z.}~\bibnamefont {Chen}}, \bibinfo {author} {\bibfnamefont {A.}~\bibnamefont
  {Megrant}}, \bibinfo {author} {\bibfnamefont {R.}~\bibnamefont {Barends}},
  \bibinfo {author} {\bibfnamefont {B.}~\bibnamefont {Campbell}}, \bibinfo
  {author} {\bibfnamefont {B.}~\bibnamefont {Chiaro}}, \bibinfo {author}
  {\bibfnamefont {A.}~\bibnamefont {Dunsworth}}, \bibinfo {author}
  {\bibfnamefont {E.}~\bibnamefont {Jeffrey}}, \bibinfo {author} {\bibfnamefont
  {J.}~\bibnamefont {Kelly}}, \bibinfo {author} {\bibfnamefont
  {J.}~\bibnamefont {Mutus}}, \bibinfo {author} {\bibfnamefont {P.~J.~J.}\
  \bibnamefont {O'Malley}}, \bibinfo {author} {\bibfnamefont {C.}~\bibnamefont
  {Quintana}}, \bibinfo {author} {\bibfnamefont {D.}~\bibnamefont {Sank}},
  \bibinfo {author} {\bibfnamefont {A.}~\bibnamefont {Vainsencher}}, \bibinfo
  {author} {\bibfnamefont {J.}~\bibnamefont {Wenner}}, \bibinfo {author}
  {\bibfnamefont {T.~C.}\ \bibnamefont {White}}, \bibinfo {author}
  {\bibfnamefont {A.}~\bibnamefont {Polkovnikov}}, \ and\ \bibinfo {author}
  {\bibfnamefont {J.~M.}\ \bibnamefont {Martinis}},\ }\bibfield  {title}
  {\enquote {\bibinfo {title} {Ergodic dynamics and thermalization in an
  isolated quantum system},}\ }\href {\doibase 10.1038/nphys3830} {\bibfield
  {journal} {\bibinfo  {journal} {Nature Physics}\ }\textbf {\bibinfo {volume}
  {12}},\ \bibinfo {pages} {1037} (\bibinfo {year} {2016})}\BibitemShut
  {NoStop}%
\bibitem [{\citenamefont {Ruebeck}\ \emph {et~al.}(2017)\citenamefont
  {Ruebeck}, \citenamefont {Lin},\ and\ \citenamefont
  {Pattanayak}}]{Ruebeck_EntanglementItsRelationship_2017}%
  \BibitemOpen
  \bibfield  {author} {\bibinfo {author} {\bibfnamefont {J.~B.}\ \bibnamefont
  {Ruebeck}}, \bibinfo {author} {\bibfnamefont {J.}~\bibnamefont {Lin}}, \ and\
  \bibinfo {author} {\bibfnamefont {A.~K.}\ \bibnamefont {Pattanayak}},\
  }\bibfield  {title} {\enquote {\bibinfo {title} {Entanglement and its
  relationship to classical dynamics},}\ }\href {\doibase
  10.1103/PhysRevE.95.062222} {\bibfield  {journal} {\bibinfo  {journal}
  {Physical Review E}\ }\textbf {\bibinfo {volume} {95}},\ \bibinfo {pages}
  {062222} (\bibinfo {year} {2017})}\BibitemShut {NoStop}%
\bibitem [{\citenamefont {Kumari}\ and\ \citenamefont
  {Ghose}(2018)}]{Kumari_QuantumclassicalCorrespondenceVicinity_2018}%
  \BibitemOpen
  \bibfield  {author} {\bibinfo {author} {\bibfnamefont {M.}~\bibnamefont
  {Kumari}}\ and\ \bibinfo {author} {\bibfnamefont {S.}~\bibnamefont {Ghose}},\
  }\bibfield  {title} {\enquote {\bibinfo {title} {Quantum-classical
  correspondence in the vicinity of periodic orbits},}\ }\href {\doibase
  10.1103/PhysRevE.97.052209} {\bibfield  {journal} {\bibinfo  {journal} {Phys.
  Rev. E}\ }\textbf {\bibinfo {volume} {97}},\ \bibinfo {pages} {052209}
  (\bibinfo {year} {2018})}\BibitemShut {NoStop}%
\bibitem [{\citenamefont {Yunger~Halpern}\ \emph {et~al.}(2018)\citenamefont
  {Yunger~Halpern}, \citenamefont {Swingle},\ and\ \citenamefont
  {Dressel}}]{YungerHalpern_QuasiprobabilityOutofTimeOrderedCorrelator_2018}%
  \BibitemOpen
  \bibfield  {author} {\bibinfo {author} {\bibfnamefont {N.}~\bibnamefont
  {Yunger~Halpern}}, \bibinfo {author} {\bibfnamefont {B.}~\bibnamefont
  {Swingle}}, \ and\ \bibinfo {author} {\bibfnamefont {J.}~\bibnamefont
  {Dressel}},\ }\bibfield  {title} {\enquote {\bibinfo {title}
  {Quasiprobability {{Behind}} the {{Out}}-of-{{Time}}-{{Ordered
  Correlator}}},}\ }\href {\doibase 10.1103/PhysRevA.97.042105} {\bibfield
  {journal} {\bibinfo  {journal} {Physical Review A}\ }\textbf {\bibinfo
  {volume} {97}},\ \bibinfo {pages} {042105} (\bibinfo {year}
  {2018})}\BibitemShut {NoStop}%
\bibitem [{\citenamefont {Tsuji}\ \emph {et~al.}(2018)\citenamefont {Tsuji},
  \citenamefont {Shitara},\ and\ \citenamefont
  {Ueda}}]{Tsuji_OutoftimeorderFluctuationdissipationTheorem_2018}%
  \BibitemOpen
  \bibfield  {author} {\bibinfo {author} {\bibfnamefont {N.}~\bibnamefont
  {Tsuji}}, \bibinfo {author} {\bibfnamefont {T.}~\bibnamefont {Shitara}}, \
  and\ \bibinfo {author} {\bibfnamefont {M.}~\bibnamefont {Ueda}},\ }\bibfield
  {title} {\enquote {\bibinfo {title} {Out-of-time-order
  fluctuation-dissipation theorem},}\ }\href {\doibase
  10.1103/PhysRevE.97.012101} {\bibfield  {journal} {\bibinfo  {journal}
  {Physical Review E}\ }\textbf {\bibinfo {volume} {97}} (\bibinfo {year}
  {2018}),\ 10.1103/PhysRevE.97.012101},\ \Eprint
  {http://arxiv.org/abs/1612.08781} {arXiv:1612.08781} \BibitemShut {NoStop}%
\bibitem [{\citenamefont {Tsuji}\ and\ \citenamefont
  {Ueda}(2018)}]{Tsuji_FluctuationTheoremQuantumstate_2018}%
  \BibitemOpen
  \bibfield  {author} {\bibinfo {author} {\bibfnamefont {N.}~\bibnamefont
  {Tsuji}}\ and\ \bibinfo {author} {\bibfnamefont {M.}~\bibnamefont {Ueda}},\
  }\bibfield  {title} {\enquote {\bibinfo {title} {Fluctuation theorem for
  quantum-state statistics},}\ }\href@noop {} {\bibfield  {journal} {\bibinfo
  {journal} {arXiv:1807.11683 [cond-mat, physics:quant-ph]}\ } (\bibinfo {year}
  {2018})},\ \Eprint {http://arxiv.org/abs/1807.11683} {arXiv:1807.11683
  [cond-mat, physics:quant-ph]} \BibitemShut {NoStop}%
\bibitem [{\citenamefont {Zhuang}\ \emph {et~al.}(2019)\citenamefont {Zhuang},
  \citenamefont {Schuster}, \citenamefont {Yoshida},\ and\ \citenamefont
  {Yao}}]{Zhuang_ScramblingComplexityPhase_2019}%
  \BibitemOpen
  \bibfield  {author} {\bibinfo {author} {\bibfnamefont {Q.}~\bibnamefont
  {Zhuang}}, \bibinfo {author} {\bibfnamefont {T.}~\bibnamefont {Schuster}},
  \bibinfo {author} {\bibfnamefont {B.}~\bibnamefont {Yoshida}}, \ and\
  \bibinfo {author} {\bibfnamefont {N.~Y.}\ \bibnamefont {Yao}},\ }\bibfield
  {title} {{\selectlanguage {en}\enquote {\bibinfo {title} {Scrambling and
  complexity in phase space},}\ }}\href {\doibase 10.1103/PhysRevA.99.062334}
  {\bibfield  {journal} {\bibinfo  {journal} {Physical Review A}\ }\textbf
  {\bibinfo {volume} {99}},\ \bibinfo {pages} {062334} (\bibinfo {year}
  {2019})}\BibitemShut {NoStop}%
\bibitem [{\citenamefont {{Mu{\~n}oz-Arias}}\ \emph {et~al.}(2021)\citenamefont
  {{Mu{\~n}oz-Arias}}, \citenamefont {Poggi},\ and\ \citenamefont
  {Deutsch}}]{Munoz-Arias_NonlinearDynamicsQuantum_2021}%
  \BibitemOpen
  \bibfield  {author} {\bibinfo {author} {\bibfnamefont {M.~H.}\ \bibnamefont
  {{Mu{\~n}oz-Arias}}}, \bibinfo {author} {\bibfnamefont {P.~M.}\ \bibnamefont
  {Poggi}}, \ and\ \bibinfo {author} {\bibfnamefont {I.~H.}\ \bibnamefont
  {Deutsch}},\ }\bibfield  {title} {\enquote {\bibinfo {title} {Nonlinear
  dynamics and quantum chaos of a family of kicked \$p\$-spin models},}\ }\href
  {\doibase 10.1103/PhysRevE.103.052212} {\bibfield  {journal} {\bibinfo
  {journal} {Physical Review E}\ }\textbf {\bibinfo {volume} {103}},\ \bibinfo
  {pages} {052212} (\bibinfo {year} {2021})}\BibitemShut {NoStop}%
\bibitem [{\citenamefont {Lerose}\ and\ \citenamefont
  {Pappalardi}(2020{\natexlab{b}})}]{Lerose_OriginSlowGrowth_2018}%
  \BibitemOpen
  \bibfield  {author} {\bibinfo {author} {\bibfnamefont {A.}~\bibnamefont
  {Lerose}}\ and\ \bibinfo {author} {\bibfnamefont {S.}~\bibnamefont
  {Pappalardi}},\ }\bibfield  {title} {\enquote {\bibinfo {title} {Origin of
  the slow growth of entanglement entropy in long-range interacting spin
  systems},}\ }\href {\doibase 10.1103/PhysRevResearch.2.012041} {\bibfield
  {journal} {\bibinfo  {journal} {Phys. Rev. Research}\ }\textbf {\bibinfo
  {volume} {2}},\ \bibinfo {pages} {012041} (\bibinfo {year}
  {2020}{\natexlab{b}})}\BibitemShut {NoStop}%
\bibitem [{\citenamefont {Sinha}\ \emph {et~al.}(2021)\citenamefont {Sinha},
  \citenamefont {Ray},\ and\ \citenamefont {Sinha}}]{Sinha2021_fingerprint}%
  \BibitemOpen
  \bibfield  {author} {\bibinfo {author} {\bibfnamefont {S.}~\bibnamefont
  {Sinha}}, \bibinfo {author} {\bibfnamefont {S.}~\bibnamefont {Ray}}, \ and\
  \bibinfo {author} {\bibfnamefont {S.}~\bibnamefont {Sinha}},\ }\bibfield
  {title} {\enquote {\bibinfo {title} {Fingerprint of chaos and quantum scars
  in kicked {D}icke model: an out-of-time-order correlator study},}\ }\href
  {https://iopscience.iop.org/article/10.1088/1361-648X/abe26b} {\bibfield
  {journal} {\bibinfo  {journal} {Journal of Physics: Condensed Matter}\
  }\textbf {\bibinfo {volume} {33}},\ \bibinfo {pages} {174005} (\bibinfo
  {year} {2021})}\BibitemShut {NoStop}%
\bibitem [{\citenamefont {Lin}\ \emph {et~al.}(2021)\citenamefont {Lin},
  \citenamefont {Lin}, \citenamefont {Ku}, \citenamefont {Lambert},
  \citenamefont {Chen},\ and\ \citenamefont
  {Nori}}]{Lin_WitnessingQuantumScrambling_2020}%
  \BibitemOpen
  \bibfield  {author} {\bibinfo {author} {\bibfnamefont {J.-D.}\ \bibnamefont
  {Lin}}, \bibinfo {author} {\bibfnamefont {W.-Y.}\ \bibnamefont {Lin}},
  \bibinfo {author} {\bibfnamefont {H.-Y.}\ \bibnamefont {Ku}}, \bibinfo
  {author} {\bibfnamefont {N.}~\bibnamefont {Lambert}}, \bibinfo {author}
  {\bibfnamefont {Y.-N.}\ \bibnamefont {Chen}}, \ and\ \bibinfo {author}
  {\bibfnamefont {F.}~\bibnamefont {Nori}},\ }\bibfield  {title} {\enquote
  {\bibinfo {title} {Quantum steering as a witness of quantum scrambling},}\
  }\href {\doibase 10.1103/PhysRevA.104.022614} {\bibfield  {journal} {\bibinfo
   {journal} {Phys. Rev. A}\ }\textbf {\bibinfo {volume} {104}},\ \bibinfo
  {pages} {022614} (\bibinfo {year} {2021})}\BibitemShut {NoStop}%
\end{thebibliography}%
\end{document}